\documentclass[12,preprint]{aastex}

\usepackage{amsmath}
\usepackage{gensymb}
\usepackage{color}

 \newcommand{\bra}[1]{\left( #1 \right)}
 \newcommand{\sqb}[1]{\left[ #1 \right]}

\DeclareMathOperator\erf{erf}

\shorttitle{Upper Limit on the Milky Way Mass from the Orbit of the Sagittarius Dwarf Satellite}
\shortauthors{Dierickx \& Loeb}

\begin{document}

\title{Upper Limit on the Milky Way Mass from the Orbit of the Sagittarius Dwarf Satellite}
\author{Marion I. P. Dierickx\altaffilmark{1} and Abraham Loeb\altaffilmark{1}}
\altaffiltext{1}{Astronomy Department, Harvard University, 60 Garden Street, Cambridge, MA 02138, USA; mdierickx@cfa.harvard.edu, aloeb@cfa.harvard.edu}

\begin{abstract}
As one of the most massive Milky Way satellites, the Sagittarius dwarf galaxy has played an important role in shaping the Galactic disk and stellar halo morphologies. The disruption of Sagittarius over several close-in passages has populated the halo of our Galaxy with large-scale tidal streams and offers a unique diagnostic tool for measuring its gravitational potential. Here we test different progenitor mass models for the Milky Way and Sagittarius by modeling the full infall of the satellite. We constrain the mass of the Galaxy based on the kinematics of the satellite remnant and multiple tidal streams of Sagittarius. Our semianalytic modeling of the orbital dynamics agrees with full $N$-body simulations, and favors low values for the Milky Way mass, $\lesssim 10^{12}M_\odot$. This conclusion eases the tension between $\Lambda$CDM and the observed parameters of the Milky Way satellites.
\end{abstract}

\section{Introduction}
\label{sec:intro}

As one of our Galaxy's closest companions, the Sagittarius (Sgr) dwarf galaxy has attracted much interest since its discovery by \citet{ibata94}. Sgr is currently being accreted by the Milky Way (MW) and has experienced several passages through the outskirts of the Galactic disk \citep[e.g.][]{purcell11}. Billions of years spent in the tidal field of the MW have resulted in a prominent stream of stars stripped from the Sgr dwarf. The stream has been detected with tracer stars such as M giants from the Two Micron All Sky Survey (2MASS) \citep{ibata02, majewski03, majewski04}, as well as data from the Sloan Digital Sky Survey (SDSS) Data Release 5 \citep{belokurov06} and 8 \citep{belokurov14}. Much effort has been dedicated to mapping these remnants in six dimensions in order to exploit them as sensitive probes of the underlying MW mass distribution. The Sgr debris are especially valuable because detected stars reach large galactocentric radii of up to $\sim$100~kpc \citep{belokurov14}. In the outer regions of the Galaxy, dark matter dominates over baryonic components in shaping the gravitational potential. As some of the few luminous tracers existing at large radii, Sgr stars can therefore serve as a probe of the dark matter halo profile. 

Many studies have attempted to measure the mass, shape, and orientation of the halo based on models of the Sgr stream. Interestingly, despite the fact that the MW is the closest galactic halo available for study, constraining these parameters has proven to be highly challenging. Studies based on the Sgr debris trail have lead to ambiguous results, with evidence pointing to a range of contradictory MW halo shapes \citep{helmi04}, namely oblate \citep{johnston05}, spherical \citep{ibata01, fellhauer06}, or triaxial \citep{law09}. Similarly, a wide range of methods have been applied to measure the Galactic virial mass. The resulting estimates vary by factors of 2-3, including values as high as  $1.9^{+3.6}_{-1.7} \times 10^{12}$~M$_\odot$ \citep[although note the large uncertainty;][]{wilkinson99} and as low as $M_{200} = 5.6 \pm 1.2 \times 10^{11}$~M$_\odot$ \citep{gibbons14}. In addition to challenging MW rotation curve measurements, the scarcity of luminous tracers at large radii and the difficulty associated with measuring transverse velocities for distant stars are the main factors impeding these estimates. At large distances, the lack of proper motion measurements and small sample size complicate estimates relying on dwarf companions of the MW \citep[e.g.][]{watkins10}. Sgr provides a unique opportunity for getting better measurements of the MW mass, since the remnant core is located nearby \citep[$d\sim25$~kpc;][]{kunder09}, yet the stellar debris delineate its past orbit extending past 100~kpc \citep{belokurov14, dierickx17}.

In \citet[][hereafter Paper I]{dierickx17} we presented a new model for the orbit of Sgr over the past 8~Gyr. Growing evidence in favor of a more massive Sgr progenitor \citep{niederste10, conroy09, behroozi10, gibbons17} suggests that Sgr must have formed on the fringes of the MW halo, and sunk to the center under the effect of dynamical friction. The model aims to account for the higher progenitor mass and initial separation, therefore simulating the full infall trajectory of Sgr since its crossing of the MW virial radius at $z\sim1$. For a present-day fiducial MW mass of $\sim10^{12}$~M$_\odot$, the MW progenitor would have had a virial radius of $\sim125$~kpc at that time, with a mass $M(<125~\text{kpc}) \simeq 5-7\times10^{11}$~M$_\odot$. Initializing the position of Sgr at that time and distance, the orbital model is therefore sensitive to the amount of mass inside the starting radius.

In this study we generalize the framework of Paper I to investigate a range of possible values for the MW and Sgr virial masses. In \S~\ref{sec:light_Sgr} and \ref{sec:massive_Sgr}, we examine two possible cases for the Sgr progenitor: a low-mass case with $M_\text{halo} = 10^{10}$~M$_\odot$, based on the estimates of \citet{niederste10}; and a massive Sgr case where $M_\text{halo} = 6\times10^{10}$~M$_\odot$, based on the work of \citet{gibbons17}. In both cases we expand upon the parameter search of Paper I and now explore a range of different MW masses. These methods are described in \S~\ref{sec:SAM_methods}. Attempting to meet observed constraints on Sgr, we examine whether the present-day six-dimensional phase space coordinates of the remnant pose a limit on the mass of the MW. In \S~\ref{subsec:light_param_exploration}, we find that the mass inside the starting radius cannot exceed $\sim 9\times10^{11}$~M$_\odot$ in order for consistent Sgr orbits to exist. Extrapolating the mass profile to 200~kpc, this suggests an upper limit of $\sim1.1 \times10^{12}$~M$_\odot$ for the MW virial mass. In \S~\ref{subsec:analytic}, we propose a simple analytic explanation for this upper bound. In the framework of our Sgr orbital model, conservation of energy and the present-day galactocentric distance and velocity magnitude of Sgr together constrain the depth of the MW potential well. We show that the Sgr remnant distance and velocity at pericenter are only mutually consistent for lower MW masses. In \S~ \ref{subsec:gadget} we compare the output from our semianalytic model to equivalent full N-body simulations and find good agreement. In \S~\ref{sec:massive_Sgr}, we turn our attention to the massive Sgr case. Examining both a `slow sinking' and a `rapid sinking' scenario, we find a similar preference for a lower-mass MW host. Finally, we summarize our main conclusions in \S~\ref{sec:conclusions}.

\section{Semianalytic Model}
\label{sec:SAM_methods}

\begin{table}[bt!]
\caption{Parameters of different semianalytic runs.}
\begin{tabular}{ccccc}
\hline
NFW $M_\text{vir}$ & Hernquist $M_\text{tot}$ & Hernquist $r_\text{scale}$ & $d_\text{init}$, $z\sim1$ & $d_\text{init}$, $z\sim0.4$  \\
\hline
$6.0\times10^{11}$~M$_\odot$ & $7.5\times10^{11}$~M$_\odot$ & 32.34 kpc & 105 kpc & 142 kpc\\
$8.0\times10^{11}$~M$_\odot$ & $1.0\times10^{12}$~M$_\odot$  & 35.60 kpc & 115 kpc & 156 kpc\\
$1.0\times10^{12}$~M$_\odot$ & $1.25\times10^{11}$~M$_\odot$  & 38.34 kpc & 125 kpc & 169 kpc\\
$1.2\times10^{12}$~M$_\odot$ & $1.5\times10^{11}$~M$_\odot$  & 40.74 kpc & 132 kpc & 179 kpc\\
$1.4\times10^{12}$~M$_\odot$ & $1.75\times10^{11}$~M$_\odot$  & 42.89 kpc & 140 kpc & 189 kpc\\
\hline
\hline
\end{tabular}
\tablecomments{The first column gives the five different values of NFW virial mass tested for the MW. In all cases the MW concentration parameter is kept constant at a value of $c=10$. The second and third columns give the parameters of the corresponding Hernquist profiles. The two rightmost columns provide the starting separation $d_\text{init}$ for the `slow sinking' ($z\sim1$) and `rapid sinking' ($z\sim0.4$) cases. 
\label{tab:params}}
\end{table}

Given the current three-dimensional position and velocity of a test particle in a known external MW gravitational potential field, the orbit can in principle be integrated backwards simply by reversing time in the equations of motion. This technique has been widely applied in order to delineate the past trajectories of MW satellites, including Sgr \citep[e.g.][]{law10, veraciro13} and the Large Magellanic Cloud (LMC) \citep[e.g.][]{besla07, kallivayalil13}. However, this method does not capture non time-reversible effects such as dynamical friction and tidal stripping of the satellite. With evidence pointing to Sgr masses as high as $6\times10^{10}$~M$_\odot$ \citep{gibbons17}, dynamical friction is expected to play an important role in reducing orbital energy and causing Sgr to sink in towards smaller Galactocentric distances \citep{jiang00}. Tidal effects are evident not only from the large-scale stream of stripped stars described in \S~\ref{sec:intro}, but also from the fact that the galaxy appears elongated toward the plane of the MW disk \citep{ibata94}. As a result, in Paper I we developed an orbital model that captures both tidal stripping and dynamical friction by integrating the equations of motion forward in time and varying the initial angular momentum of Sgr. Here we extend our analysis and perform four additional calculations, varying the MW mass from 0.6 to 1.4 times the fiducial value of $10^{12}$~M$_\odot$ \citep{xue08}. Throughout the paper, these different cases are labelled by the dimensionless parameter $0.6 \leq m \leq 1.4$.

We adopt the same formalism as in Paper I throughout the computation. The MW is described by a 3-component gravitational potential consisting of a Hernquist dark matter halo, an exponential disk, and a Hernquist bulge \citep{hernquist90}.
The MW parameter values are adapted from \citet{gomez15}. We use disk and bulge masses of 0.065$M_\text{halo}$ and 0.01$M_\text{halo}$, respectively. Five different values of the Hernquist halo mass $M_\text{halo}$ and scale radius $r_\text{H}$ are explored, as summarized in Table~\ref{tab:params}. We choose a disk scale radius $b_\text{0}=3.5$~kpc and a bulge scale radius of $c_\text{0} = 0.15 b_\text{0}$. The Sgr progenitor is kept fixed and simply modeled by a Hernquist dark matter halo. For both galaxies the Hernquist total halo mass and scale radius are tuned to match the enclosed mass of Navarro, Frenk and White profiles \citep[NFW;][]{navarro97} used in the literature (for more details, see Paper I). In \S~\ref{sec:light_Sgr} we explore parameters for a low-mass Sgr progenitor with $M_\text{halo} = 10^{10}$~M$_\odot$ and concentration parameter $c=8$, following the mass estimates by \citet{niederste10}. In \S~\ref{sec:massive_Sgr} we explore the massive Sgr progenitor case, with a halo mass of $6\times10^{10}$~M$_\odot$, following the recent estimates by \citet{gibbons17}.

The initial angular momentum of the Sgr progenitor is described by two parameters: $v_{\text{init}}$, the magnitude of the Sgr velocity, and $\theta_{\text{init}}$, the angle between the velocity vector and the direction to the MW center. We integrate a $30\times30$ grid of 900 orbits, with $\theta_{\text{init}}$ ranging from $10^\circ$ to $90^\circ$, and $v_{\text{init}}$ ranging from 0 to the NFW escape velocity for the MW potential and Sgr starting radius in each case.

For the low-mass case, we choose a fiducial value of $z\sim1$ for the redshift at which Sgr first crossed the MW virial radius (as in Paper I). The corresponding lookback time is consistent with the age of M-giants in the stream, which has been estimated to be $8\pm1.5$~Gyr \citep{bellazzini06}. For every Sgr trajectory calculated with the semianalytic model, we extract the best-match snapshot occurring between $t=7$ and $t=9$~Gyr. We allow for $\sim8$~Gyr of orbital evolution in order to isolate pericenter passages analogous to the present day. The starting distance of the Sgr progenitor is determined by calculating the virial radius a MW precursor with roughly half of the present-day mass would have had at $z = 1$\footnote{We assume that the MW grows from the inside out \citep{loeb03} and so the region interior to the Sgr orbit is unaffected by later growth of the MW.}. For the massive Sgr progenitor case, we additionally investigate a `rapid sinking' scenario, where the virial radius crossing occurred only $\sim4$~Gyr ago. At the corresponding redshift of $z\sim0.37$, the MW has accumulated more mass and grown in size compared to $z\sim1$. Therefore, the Sgr starting distances, taken as the estimated MW virial radius at that redshift, are larger than in the $z\sim1$ case. In this late entry scenario, we allow for approximately 4~Gyr of orbital evolution and extract the best-match snapshot occurring between 3 and 5~Gyr after entry. The MW halo parameters and Sgr starting distances are summarized in Table~\ref{tab:params}. 

The Sgr satellite experiences dynamical friction and tidal stripping throughout its infall. Dynamical friction is modeled with a time-dependent modification of the Chandrasekhar formula \citep[][eq. 7.18]{binney87} tuned to provide good agreement with N-body realizations. Tidal stripping is incorporated by calculating the minimum tidal radius at each time step and neglecting the Sgr mass outside of that radius. At each time step, a sphere of possible points corresponding to the Sun's location is searched to provide a good match to the position and velocity of the Sgr remnant observed today. This is done by comparing simulated and observed 6-dimensional phase-space coordinates with a Chi-squared statistic, as in Paper I. We emphasize the fact that we are matching the coordinates of the Sgr main body, not attempting to fit properties of the tidal stream. Following \citet{veraciro13}, the observed current three-dimensional position of Sgr is determined by its galactic longitude and latitude $(l,b) = (5.6^\circ, -14.2^\circ)$ \citep{majewski03} and its heliocentric distance $d_\text{helio} = 25\pm2$~kpc \citep{kunder09}. The Sgr velocity vector is determined by the heliocentric radial velocity, measured at $140\pm0.33$~km~s$^{-1}$ \citep[weighted mean of the Sgr,N and M54 average velocities estimates in Table 5 of][]{bellazzini08}, and by its proper motion in the equatorial coordinate system: $(\mu_\alpha, \mu_\delta) = (-2.95 \pm 0.18, -1.19 \pm 0.16)$~mas~yr$^{-1}$ \citep{massari13}.

\section{Low-Mass Sgr case}
\label{sec:light_Sgr}

\subsection{Parameter Space Exploration}
\label{subsec:light_param_exploration}

\begin{figure*}[hbt!]
\begin{center}
\includegraphics[scale=0.19,trim = 3mm 0mm 1.8mm 2mm, clip]{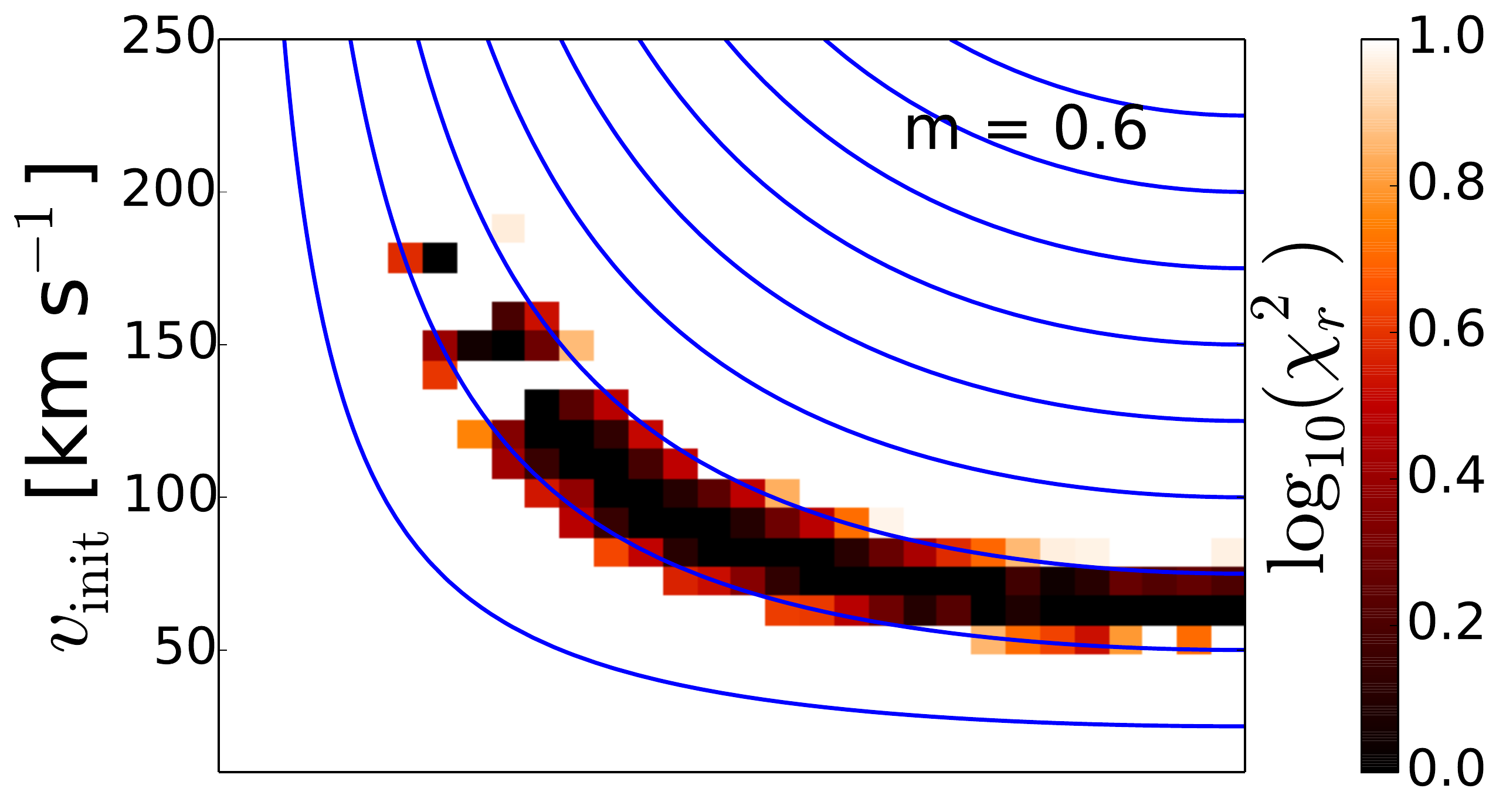} 
\includegraphics[scale=0.19,trim = 3mm 0mm 1.8mm 2mm, clip]{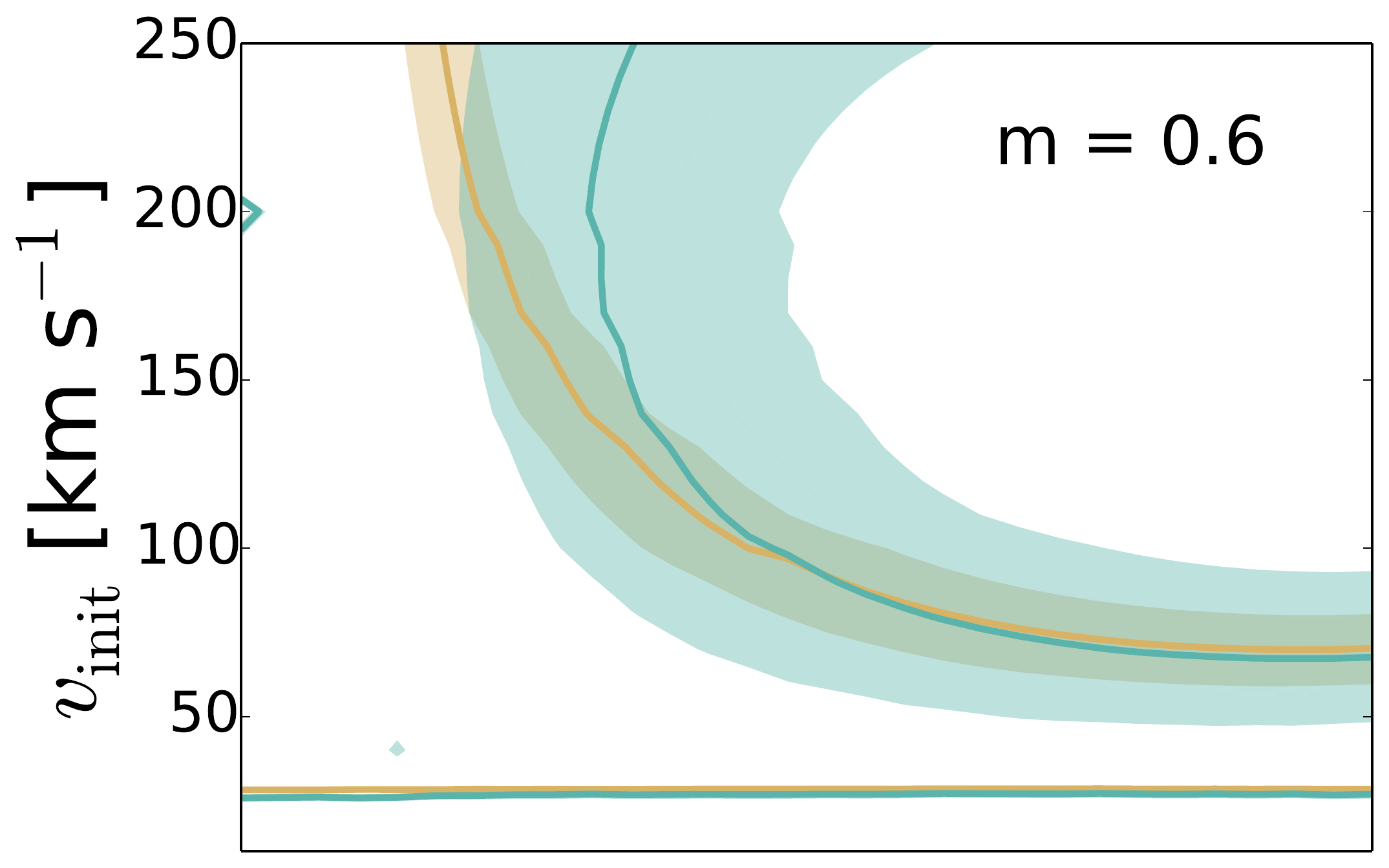} \\
\includegraphics[scale=0.19, trim = 3mm 0mm 1.8mm 2mm, clip]{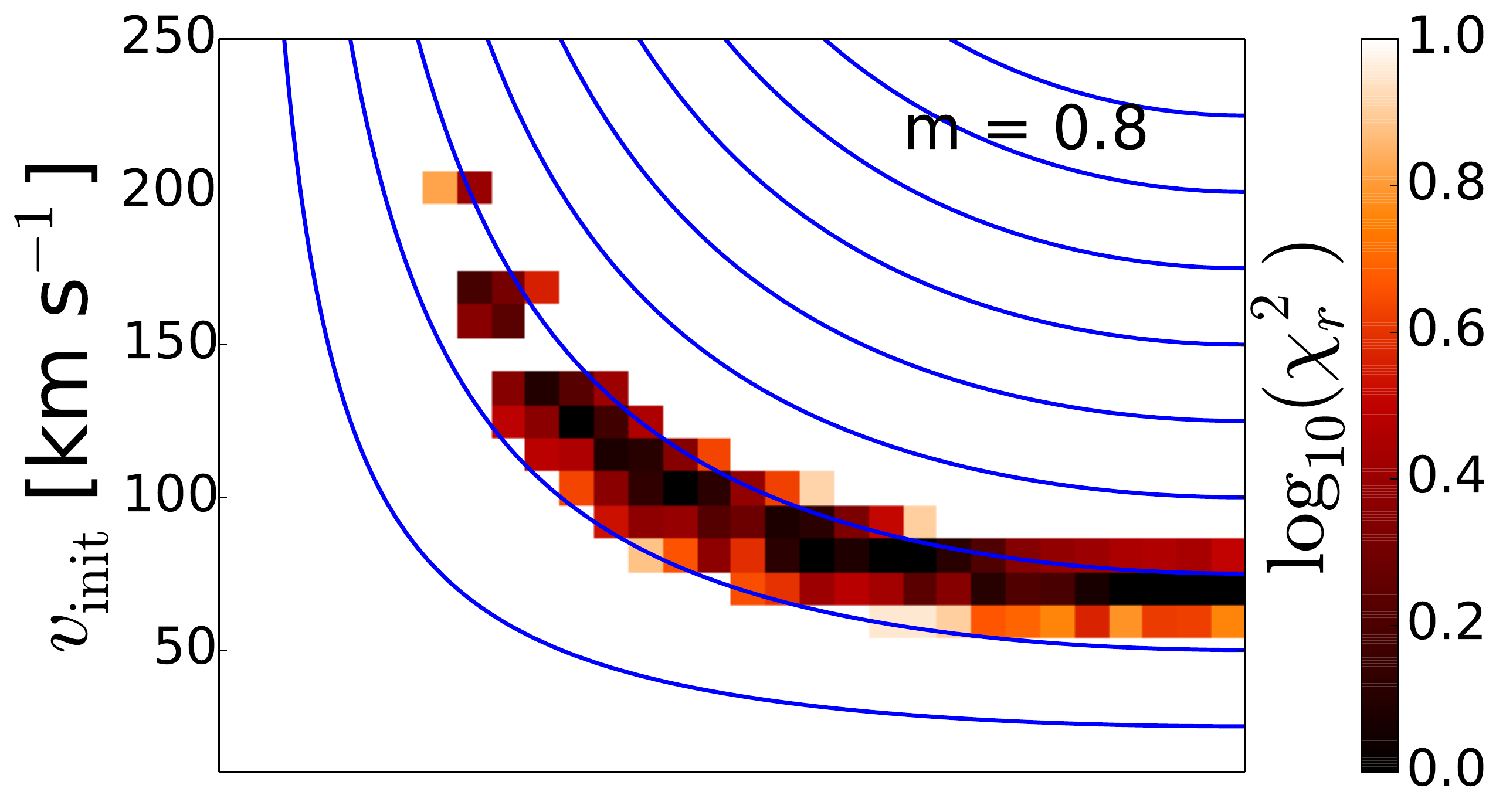} 
\includegraphics[scale=0.19, trim = 3mm 0mm 1.8mm 2mm, clip]{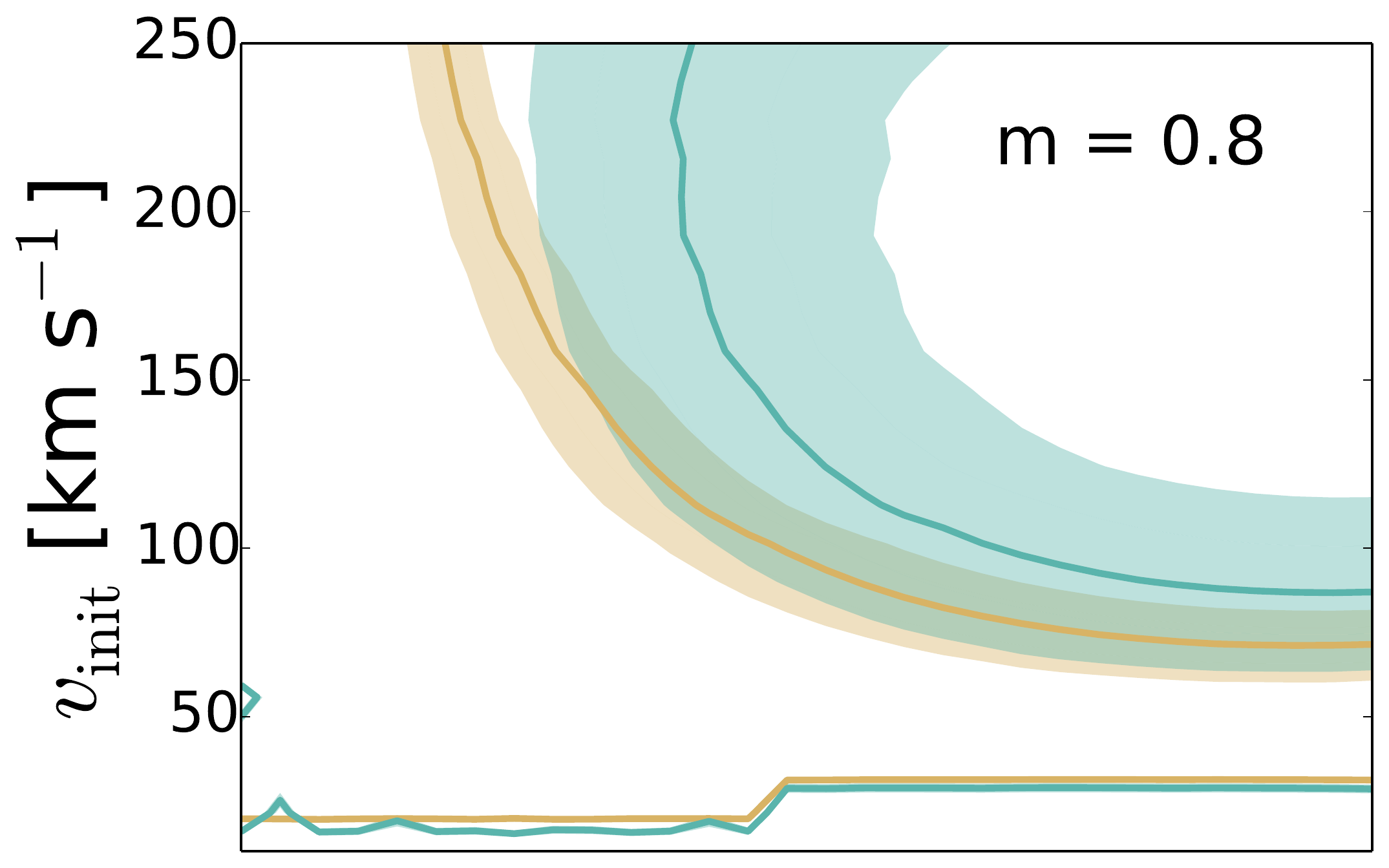} \\
\includegraphics[scale=0.19, trim = 3mm 0mm 1.8mm 2mm, clip]{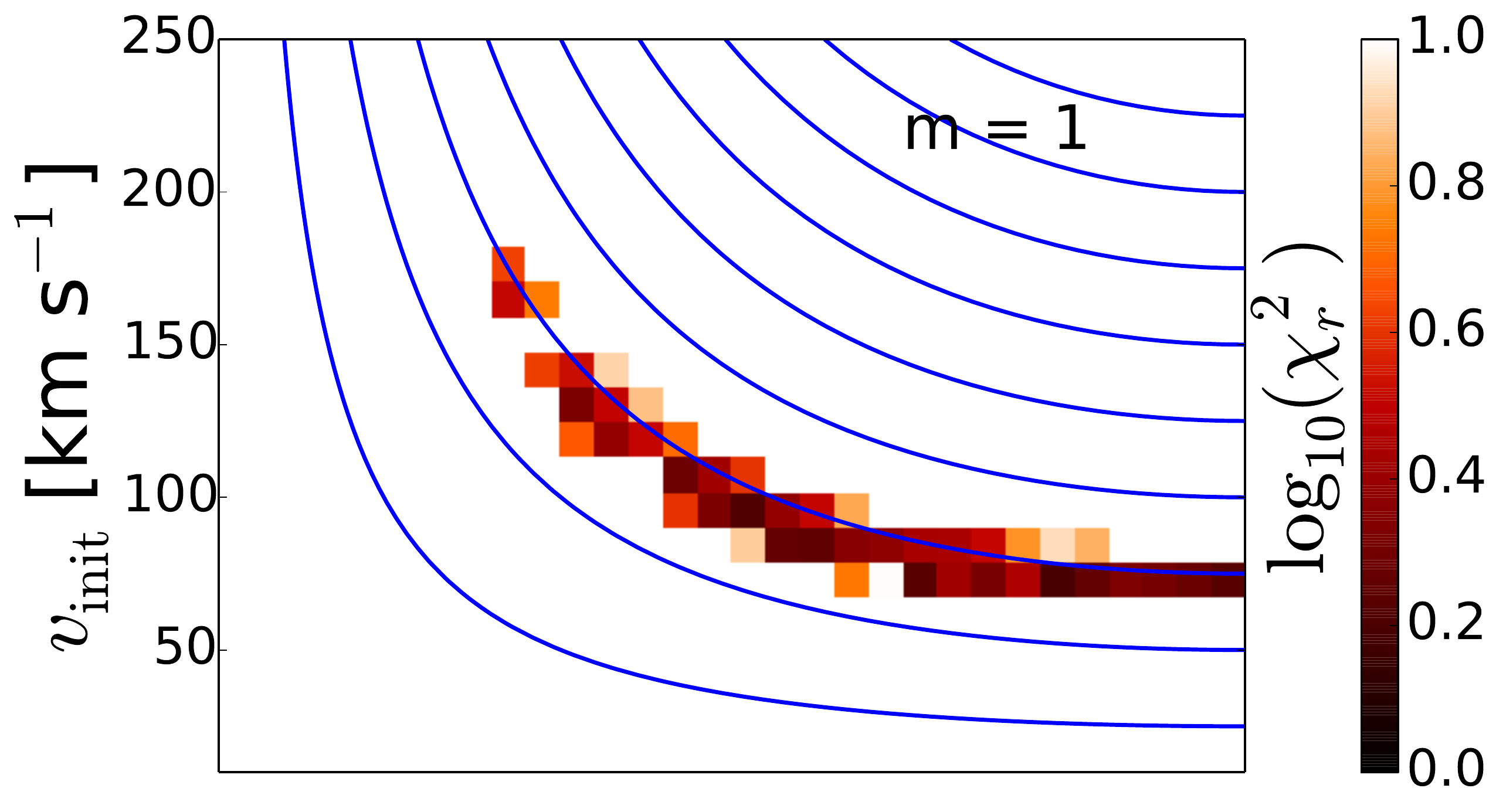} 
\includegraphics[scale=0.19, trim = 3mm 0mm 1.8mm 2mm, clip]{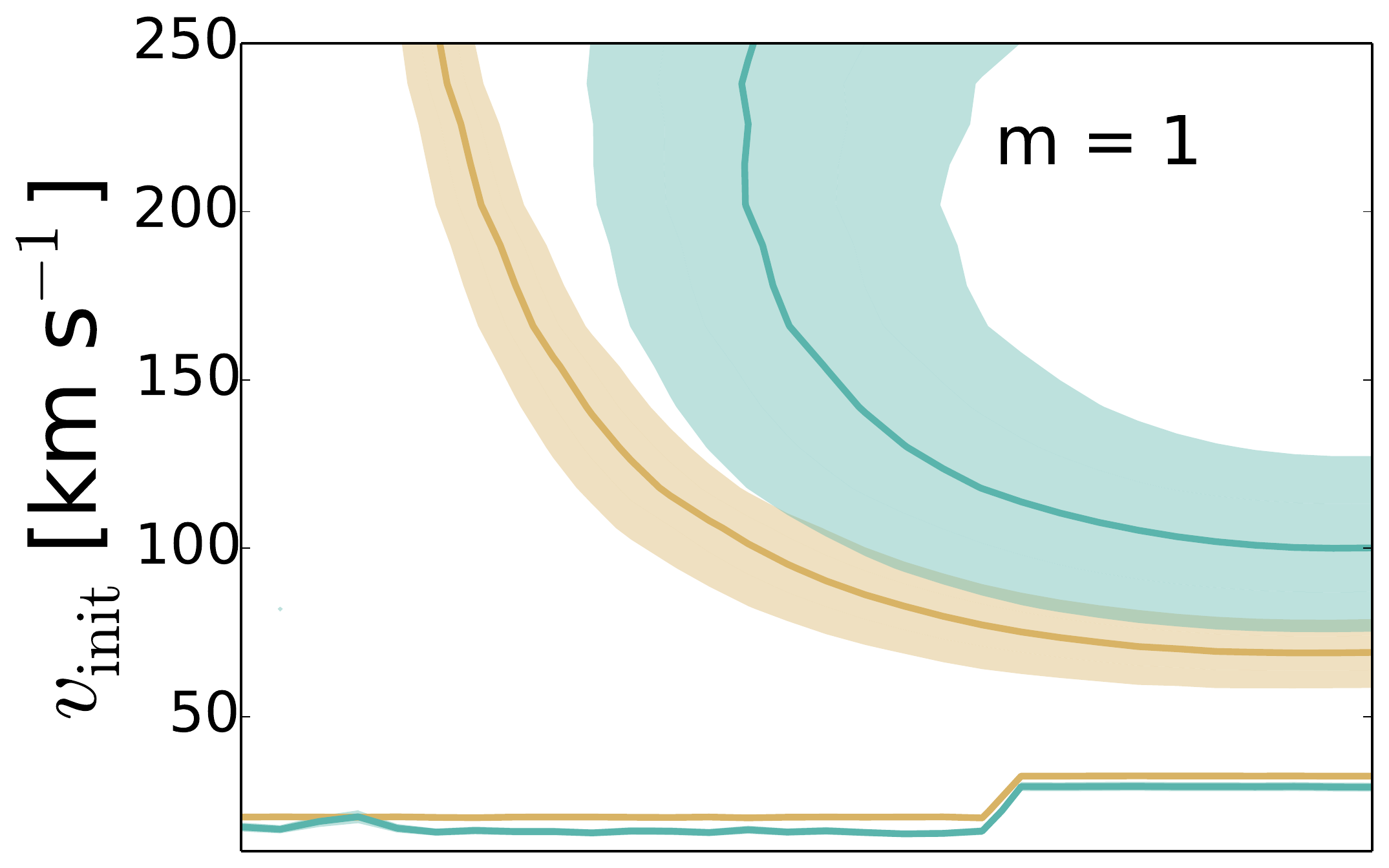} \\
\includegraphics[scale=0.19, trim = 3mm 0mm 1.8mm 2mm, clip]{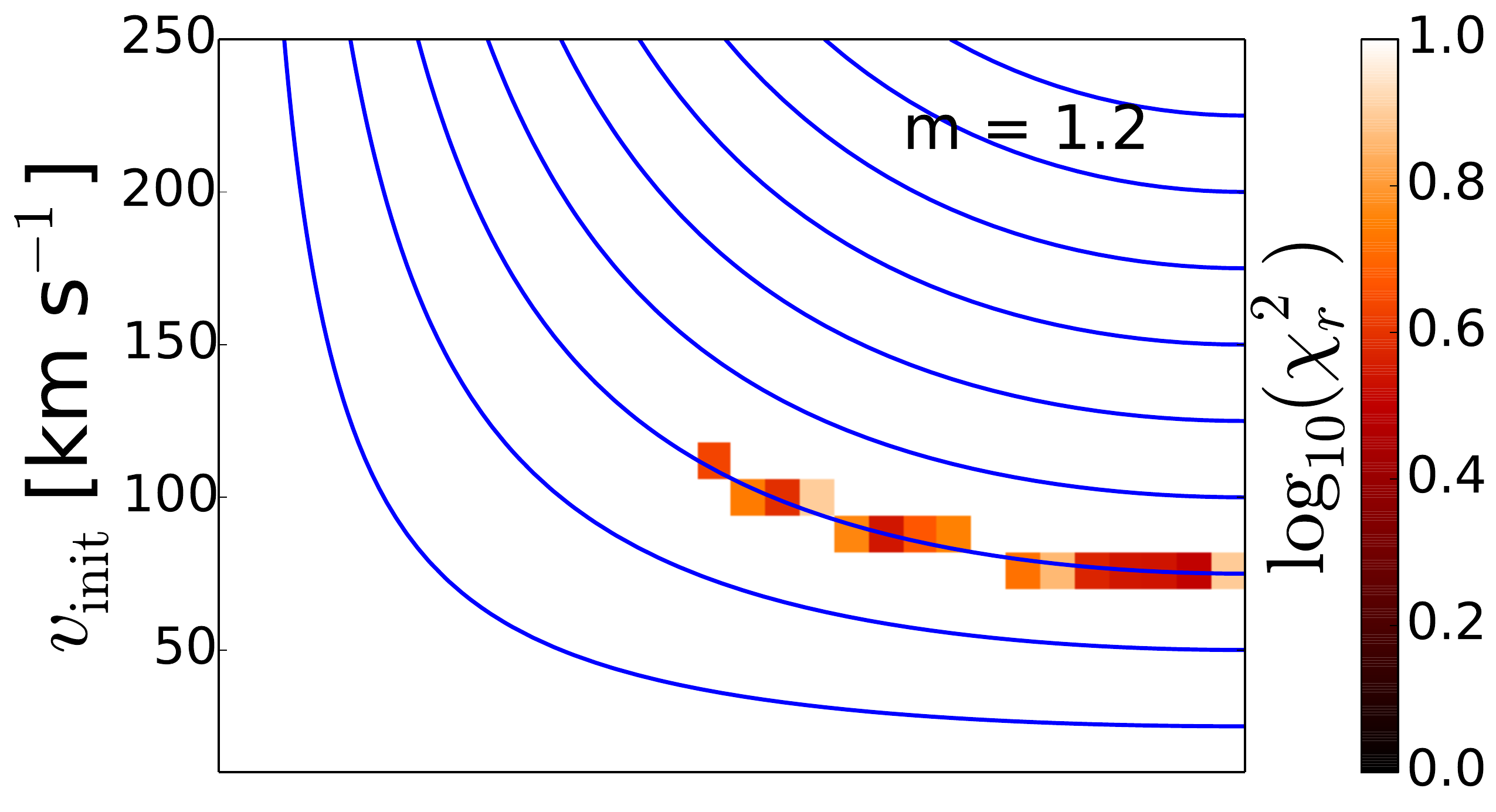} 
\includegraphics[scale=0.19, trim = 3mm 0mm 1.8mm 2mm, clip]{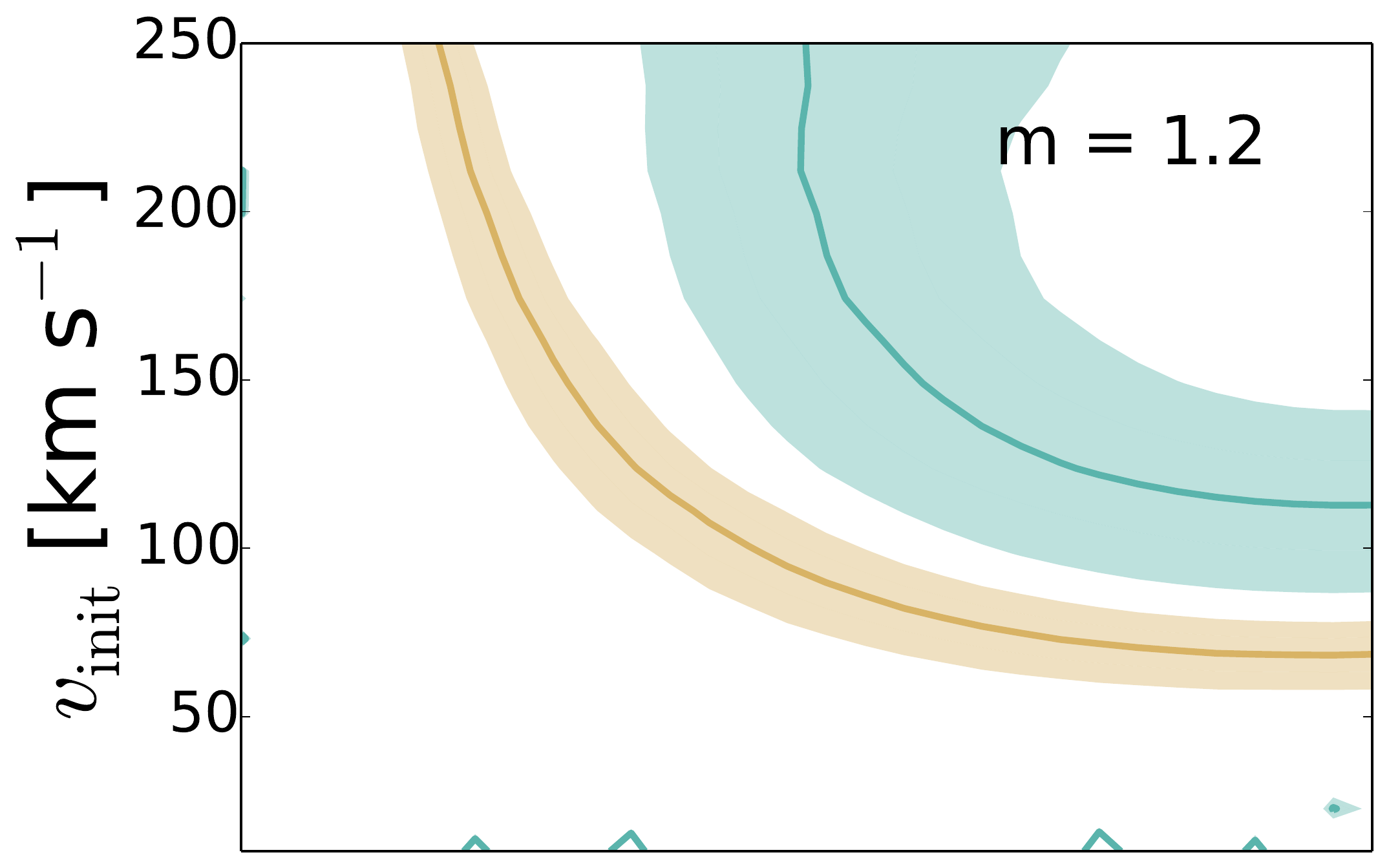} \\
\includegraphics[scale=0.19, trim = 0mm 0mm 1.8mm 2mm, clip]{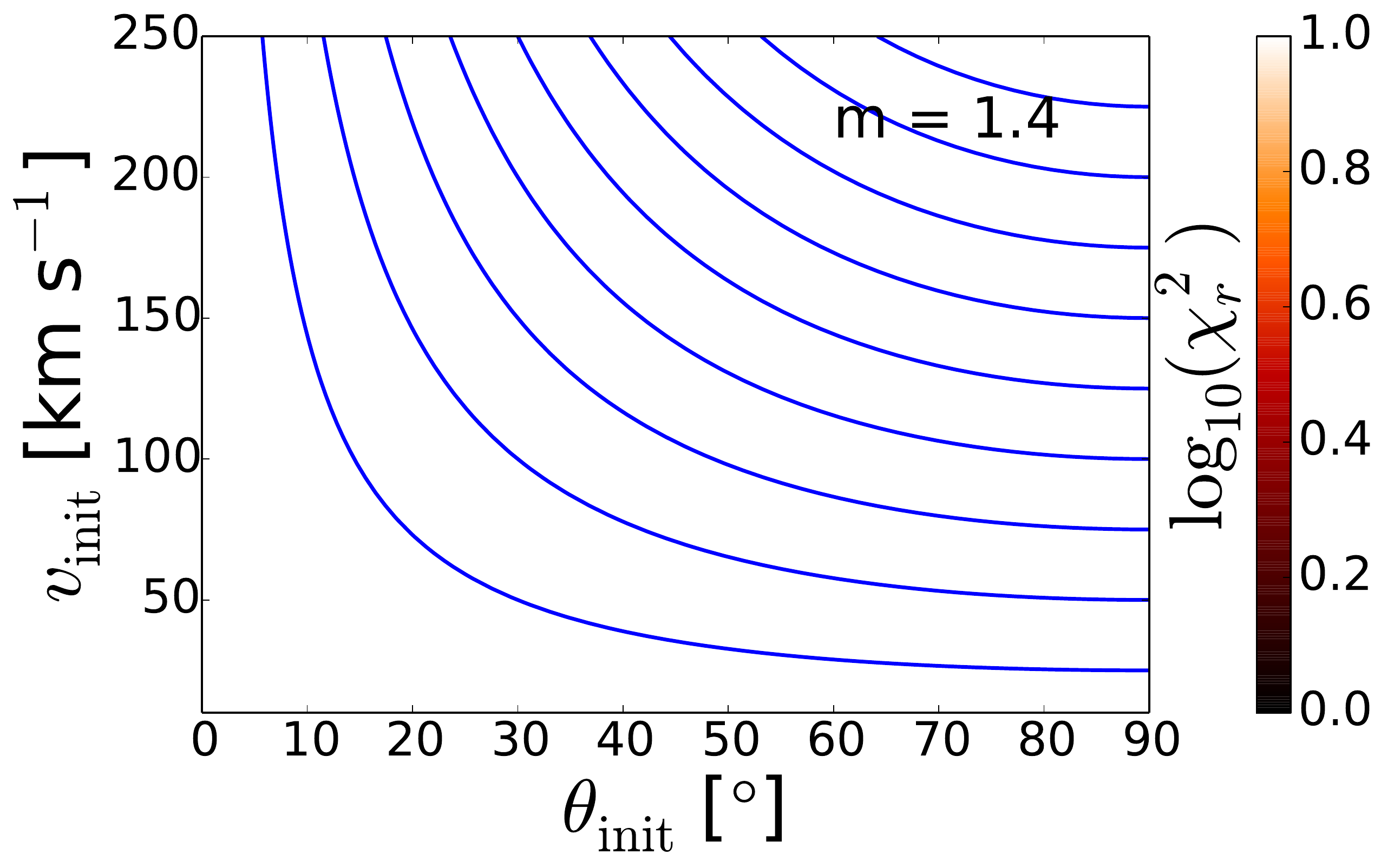} 
\includegraphics[scale=0.19, trim = 0mm 0mm 1.8mm 2mm, clip]{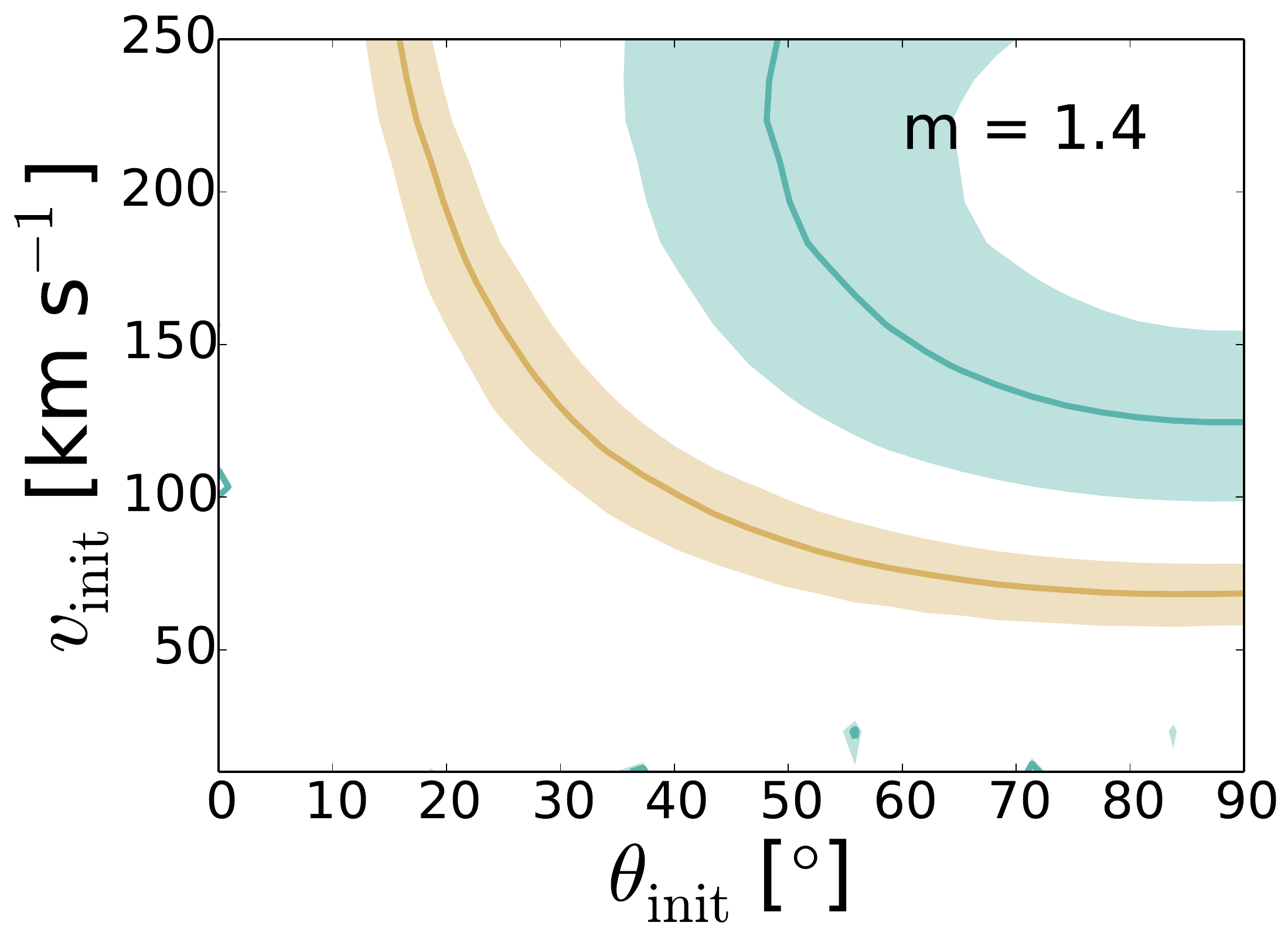} 
\caption{\textbf{Left panels:} Quality of match to the observed Sgr phase-space coordinates for different MW masses, over the initial parameter space described in \S~\ref{sec:SAM_methods}. Values of $\log_{10}(\chi^2_r) > 1$ indicate a very poor match and are uniformly colored white. Blue lines show contours of constant initial specific angular momentum per unit distance, ranging from 25 to 250~km~s$^{-1}$ for $\theta_\text{init} =90^\circ$ and are evenly spaced every 25~km~s$^{-1}$. \textbf{Right panels:} Contours of Sgr velocity magnitude (blue) and distance (brown), measured at pericenter passage, across the initial parameter space and for different MW masses. For both Galactocentric velocity and distance, the colored line maps where the observed value for Sgr today occurs in parameter space, and the shaded regions show the $2\sigma$ uncertainty interval. The lack of overlap in parameters leading to both a suitable distance and velocity grows as the Milky Way mass increases.
\label{fig:ridge1} }
\end{center}
\end{figure*}

The color maps in the left column of Figure~\ref{fig:ridge1} show regions of parameter space where agreement is found with the present-day position and velocity of Sgr, for a range of different MW masses. Darker pixels indicate lower values of the reduced chi-squared statistic calculated at the time of best match along each orbit. While initial velocities larger than 250~km~s$^{-1}$ were explored, they are not shown here as no interesting behavior is found for such high velocities.
As in Paper I, nearly radial orbits ($\theta_{\text{init}} \lesssim 30^\circ$) are excluded, while values of the initial velocity comprised between $\sim 50$ and $150$~km~s$^{-1}$ are preferred across a wide range of angles. Contours of constant angular momentum are shown as blue lines across the parameter space. For the range of MW masses investigated here, low values of $\log_{10}(\chi^2_r)$ lie on a narrow island closely aligned with the contours of constant angular momentum. This suggests that the model favors a specific value for the angular momentum the Sgr progenitor possessed upon crossing the MW's virial radius. The results of Paper I are therefore generalized to a wider range of possible MW masses beyond the chosen fiducial value. 

Two trends emerge in these color maps as the MW mass increases. First, the width of the favored region of parameter space diminishes, as fewer orbits present compatibility with present-day constraints. Moreover, at higher masses the quality of the matches declines in the remaining narrow strip of favorable parameter space. For the highest mass value considered in this study ($M_{vir} = 1.4\times10^{12}$~M$_\odot$), as defined by the criterion $\log_{10}(\chi^2_r) \leq 1$, no orbits are found to be consistent with the present-day Sgr phase-space coordinates.

The left panels of Figure~\ref{fig:ridge1} show that the present-day phase-space coordinates of Sgr are not reproduced by any combination of ($\theta_{\text{init}}$, $v_{\text{init}}$) at higher MW masses. To examine why that is the case, we consider the behavior of the satellite distance and velocity separately across parameter space. The Sgr dwarf is believed to currently be near its closest approach to the MW center. For every orbit in the grid, we extract the Galactocentric distance and velocity magnitude of the satellite at the analogous pericenter passage (the passage occurring closest to the present time). The right-hand-side panels of Figure~\ref{fig:ridge1} are contour maps of these pericenter velocities and distances. In particular, we have highlighted the contours corresponding to the observed Galactocentric radius and velocity magnitude of Sgr today (brown and blue line, respectively). The colored bands mark the $2\sigma$ interval on either side of the central contour, and therefore show which regions in ($\theta_{\text{init}}$, $v_{\text{init}}$)-space are consistent with the observed phase-space coordinates of Sgr.

At lower MW masses, the uncertainty bands and the central lines overlap, indicating the regions of parameter space consistent with both measurements. The location of the overlap zone coincides with the low $\log_{10}(\chi^2_r)$ islands described earlier for the left-hand-side color maps of Figure~\ref{fig:ridge1}. As the model MW mass increases, the divide between the contours of observed Galactocentric coordinates widens, such that there is no more overlap. While the distance contour remains stationary, the velocity band moves to the high ($\theta_{\text{init}}$, $v_{\text{init}}$) corner of parameter space. This suggests that at higher MW masses, the desired velocity is only obtained for high initial angular momentum. This high angular momentum then prevents the Sgr satellite from penetrating deep inside the halo to reach the low Galactocentric radius observed today. Conversely, at high MW masses, parameters that lead to the measured Galactocentric distance give rise to overly large velocities. 

\subsection{Conservation of Orbital Energy at Low Sgr Masses}
\label{subsec:analytic}

\begin{figure*}[hbt!]
\begin{center}
\includegraphics[scale=0.45]{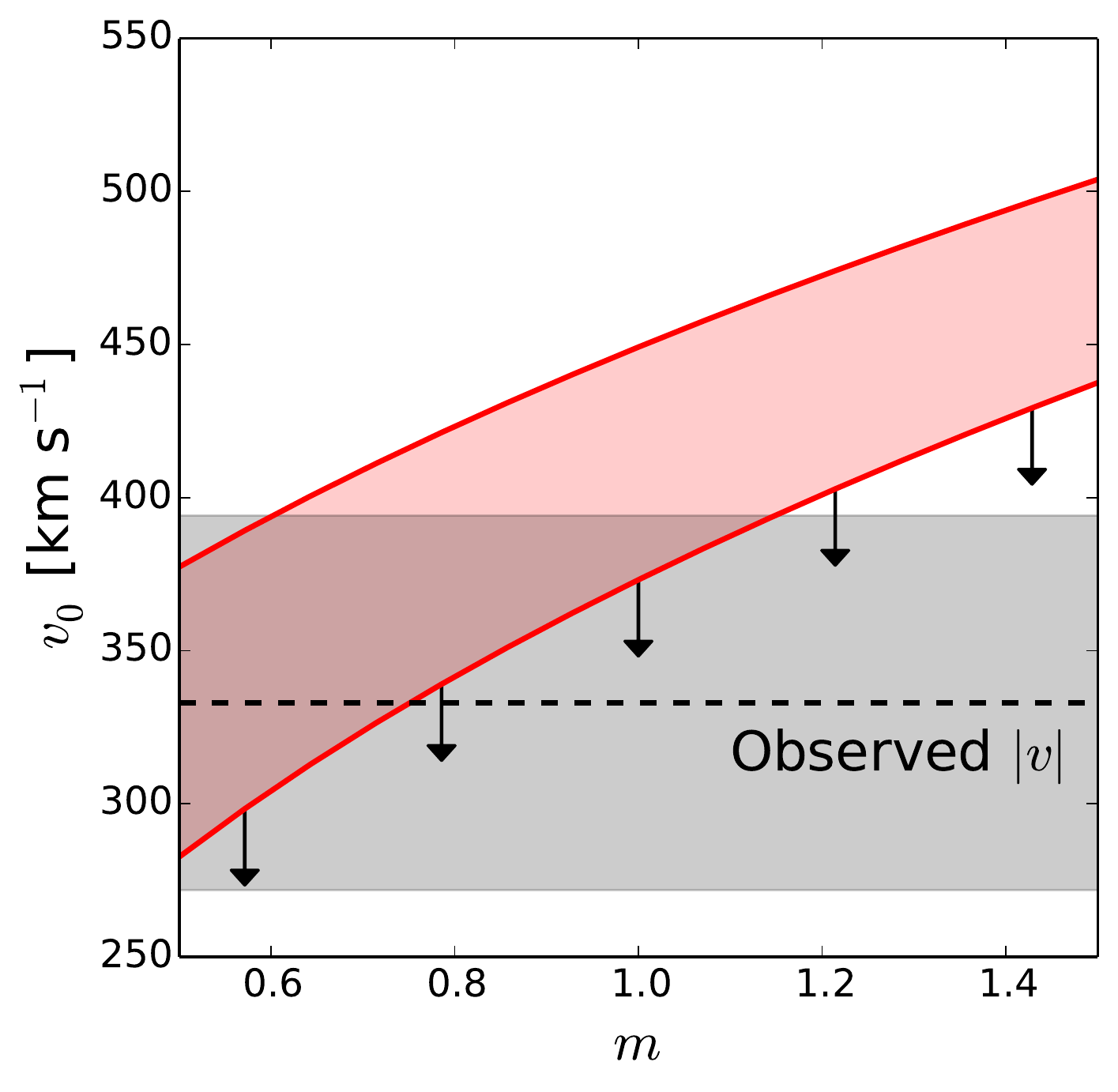} 
\caption{Analytic calculation of the Sgr velocity for different MW masses parameterized by $m$. The grey range indicates the $2\sigma$ error interval on either side of the observed total velocity magnitude for Sgr today. The two red lines delineate the range for $0 < v_{\text{init}} \lesssim 250$~km~s$^{-1}$, the maximum initial velocity considered in the parameter space shown in Figure~\ref{fig:ridge1}. The arrows indicate that the velocities shown here by the red curve are an upper bound, because dynamical friction has the effect of slowing down the satellite.
\label{fig:analytic_vel} }
\end{center}
\end{figure*}

The results highlighted in \S~\ref{subsec:light_param_exploration} show that for higher MW masses, the model has difficulty reproducing the Sgr velocity and distance simultaneously. The increasingly high Galactocentric velocity occurring when the satellite is found in the correct distance range suggests a simple interpretation based on conservation of energy at low Sgr masses. As the MW potential well deepens, the satellite gains too much velocity during infall to match the observed (relatively low) velocity today.

An upper bound on the satellite velocity can be calculated by assuming that the initial total energy equals that at Sgr's location today:
\begin{equation}
\Phi(d_\text{init}) + \frac{1}{2}v_\text{init}^2 = \Phi(d_0) + \frac{1}{2} v_0^2,
\end{equation}
\begin{equation}
\label{eq:analytic}
v_0 = \sqrt{ v_\text{init}^2 + 2\sqb{\Phi(d_\text{init}) - \Phi(d_0)  } },
\end{equation}
where $d_\text{init}$ is Sgr's starting distance (see \S~\ref{sec:SAM_methods}), $d_0\simeq 17.4$~kpc denotes the observed Galactocentric distance today, and the potential $\Phi(r)$ is that of an NFW profile with mass $M_{200c}$ and concentration $c$: 
\begin{equation}
\Phi(r) = -\frac{G M_{200c}}{r} \frac{\ln(1+cr/R_{200c})}{\ln(1+c)-c/(1+c)}.
\end{equation}
The virial radius $R_{200c}/c$ is calculated as follows:
\begin{equation}
R_{200c} = \sqb{\frac{3}{4 \pi} \frac{M_{200c}}{200 \rho_\text{c}(z) } }^{1/3},
\end{equation}
where $\rho_\text{c}(z)$ is the critical density at redshift $z$. We implement equation~\ref{eq:analytic} for different values of $M_{200c, z=0}$. In each case we make the simplifying assumptions that $M_{200c, z=1} = M_{200c, z=0}/2$ \citep[e.g.][]{torrey15,lu16}, $c_{z=1}=7$ and $c_{z=0}=10$ \citep[e.g.][]{diemer15}. The results are shown by the red band in Figure~\ref{fig:analytic_vel} for the five different MW mass models (parameterized on the $x$-axis by the multiplier $m$). The width of the band corresponds to an initial velocity ranging from zero to 250~km~s$^{-1}$, the same range considered earlier in \S~\ref{subsec:light_param_exploration}. A black dashed line marks the measured velocity magnitude of Sgr today, with $2\sigma$ uncertainty intervals shown on either side as grey regions. 

Figure~\ref{fig:analytic_vel} shows that the Sgr velocity calculated from energy conservation is only consistent with the observed range for $m \leq 1$. At higher MW masses, the velocity gained from the potential difference between the starting and final radii is too large to be consistent with the measured value of $333 \pm 30$~km~s$^{-1}$. Downward arrows illustrate the fact that the range of $v_0$ derived from equation~\ref{eq:analytic} is an upper bound, as the calculation does not include dissipative effects such as dynamical friction. Friction acts to slow down the simulated satellite and therefore contributes to bringing its velocity closer to the observed constraints. This effect is likely responsible for the marginal matches seen for $m=1.2$ in Figure~\ref{fig:ridge1}.

\begin{figure*}[hbt!]
\begin{center}
\includegraphics[scale=0.45]{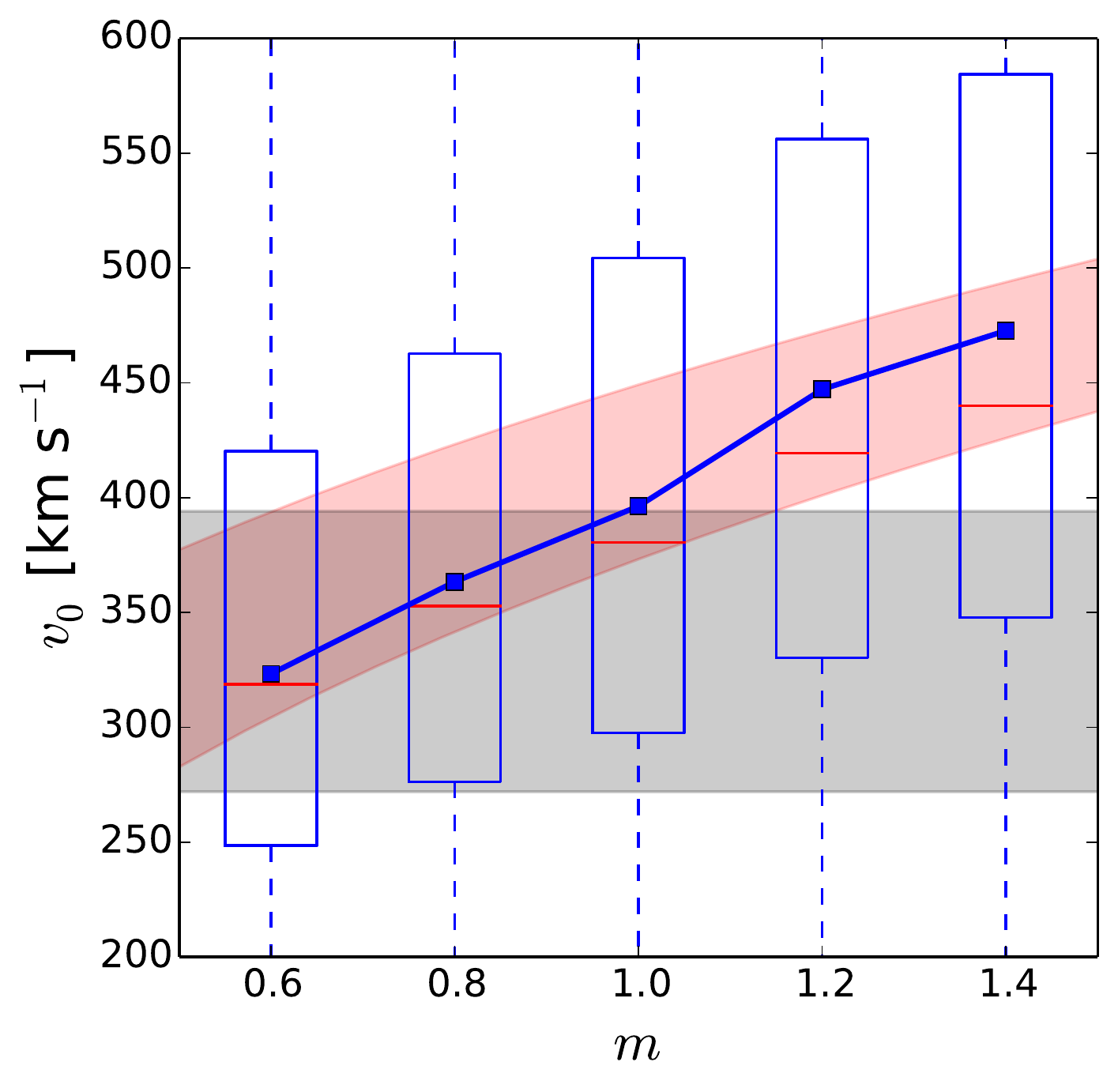} 
\caption{Comparison of analytic calculation of pericenter velocity (red range) with velocities from semianalytic models (boxes and blue line) for different MW masses. As in Figure~\ref{fig:boxplot}, the grey range indicates the $2\sigma$ error interval on either side of the observed total velocity magnitude of Sgr today. The red band delineates the range of analytically calculated pericenter velocities, for $0 < v_{\text{init}} \lesssim 250$~km~s$^{-1}$. The blue line and squares show the means of the pericenter velocities output by the point-particle model presented in \S~\ref{subsec:light_param_exploration} for different MW masses. The boxes extend from the lower to upper quartile values of the velocity data, with a red line at the median. The whiskers extend from the box to show the wide range of the data. We find that for higher MW masses, only the lowest quartile of the data is consistent with the observed velocity of Sgr.
\label{fig:boxplot} }
\end{center}
\end{figure*}

This simple conservation of energy argument offers an explanation for the present-day position and velocity of Sgr appearing only mutually consistent for lower MW masses. Next we verify whether the velocities calculated from energy conservation agree with those computed by the semianalytic model presented in \S~\ref{subsec:light_param_exploration}. The boxplots in Figure~\ref{fig:boxplot} represent the velocity distributions for the different mass bins used to test the MW potential. Because we show the results for trajectories across the full parameter space tested, there is a wide spread in Sgr velocities extending beyond the plot range. We find that most velocities within one quartile of the sample mean are approximately consistent with the energy conservation prediction of Sgr pericenter velocities. The distribution means lie fully within the predicted range. As expected, the velocities from the semianalytic model are generally lower than those calculated from conservation of energy, due to the dissipative effect of dynamical friction. As in Figure~\ref{fig:analytic_vel}, the predicted velocities rise above the observed range for MW masses above $m\sim1$. The agreement of the velocities calculated from orbital modeling with energy conservation predictions lends credence to mass constraints based on the orbital dynamics of Sgr.

\subsection{Agreement between Semianalytic and $N$-body models}
\label{subsec:gadget}

\begin{figure*}[hbt!]
\begin{center}
\includegraphics[scale=0.5]{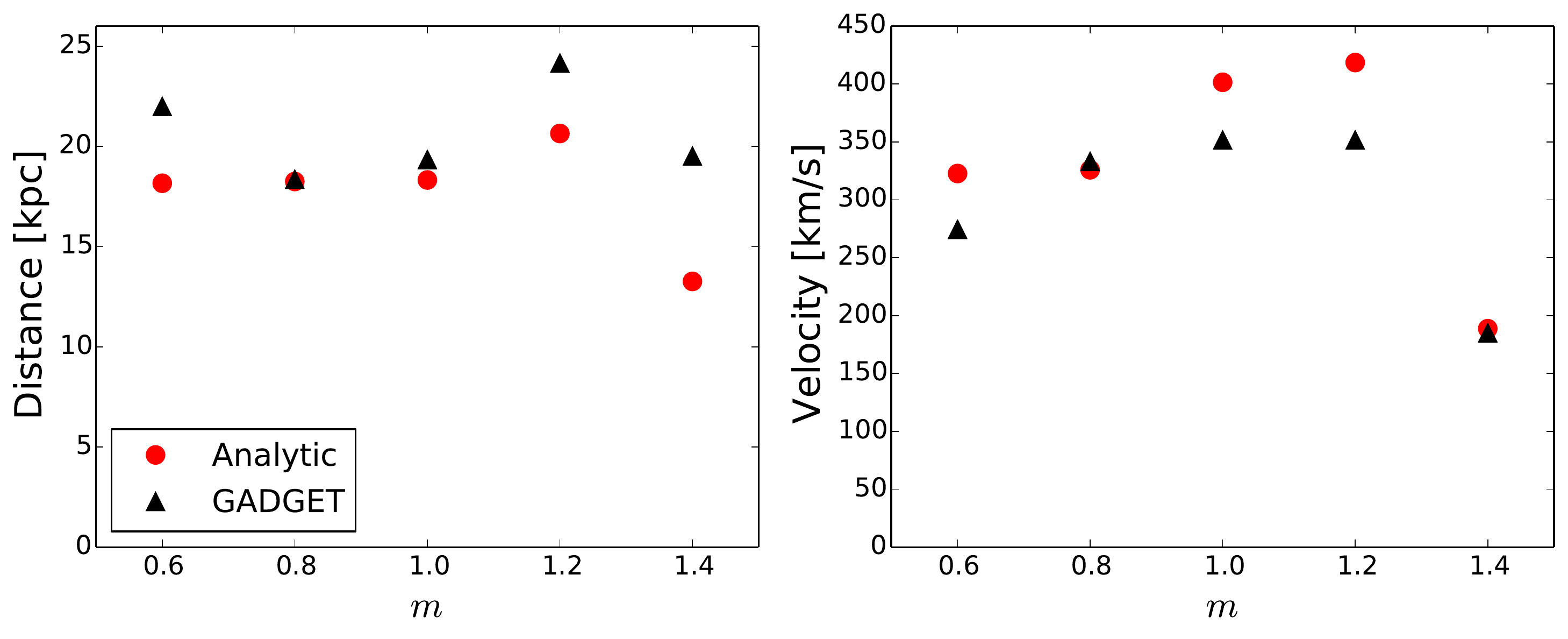} 
\caption{The galactocentric distance (left panel) and velocity magnitude (right panel) of Sgr at the time of best match in semianalytic (red circles) and GADGET (black triangles) runs, for the same initial conditions. We find good agreement between the phase-space coordinates produced by the two methods. 
\label{fig:gadget} }
\end{center}
\end{figure*}

As a further check on the analyses presented above, we seek to confirm agreement between the Sgr coordinates output by the semianalytic model and full $N$-body realizations of the same orbits. For each of the five possible MW mass values investigated here, the best-fit Sgr orbit is selected from the parameter grid presented in \S~\ref{subsec:light_param_exploration} and Figure~\ref{fig:ridge1}. Using identical initial conditions, we re-run the corresponding trajectories with the $N$-body code {\footnotesize GADGET} \citep{springel05}. We choose fixed mass resolutions of $10^6$~M$_{\odot}$ and $2\times10^5$~M$_{\odot}$ for dark matter and stellar particles, respectively. The corresponding softening lengths are 214~pc and 71~pc, respectively. We run the simulations for 9~Gyr using an adaptive time step of maximum length 10~Myr.

We extract the Sgr galactocentric distance and velocity magnitude for each simulation at the time of best match to current observables. Figure~\ref{fig:gadget} presents a comparison between the resulting quantities and the analogous phase-space coordinates produced by the semianalytic model. We find satisfactory agreement between the two methods for these five example orbits. The high MW mass case ($m$=1.4) lies outside of the distance and velocity ranges outlined by the other four data points. This is to be expected given the lack of successful orbits at such high host masses (see Figure~\ref{fig:ridge1}). Across these test simulations, the semianalytic model appears to slightly underestimate galactocentric distances and to overestimate velocity magnitudes. This is consistent with the behavior noted in Paper I and suggests stronger dynamical friction may further improve the semianalytic treatment. While in this case we only investigate the agreement between phase-space coordinates of the Sgr centroid, the distribution of particles in the best-match simulation corresponding to a MW mass of $10^{12}$~M$_{\odot}$ was studied in Paper I.

\section{Massive Sgr case}
\label{sec:massive_Sgr}

The detailed kinematic reconstruction of the Sgr tidal debris by \citet{law10} used an initial total mass of $6.4\times10^{8}$~M$_{\odot}$ for the Sgr satellite. However, several studies point to a Sgr remnant mass significantly exceeding that value. \citet{ibata97} and \citet{ibata98} estimate lower bounds of $10^9$~M$_{\odot}$ for the mass of the dwarf today. With the discovery of previously unseen branches of the stream, the total luminosity budget of the progenitor galaxy is now believed to be on the order of $10^{8}$~L$_{\odot}$ \citep{niederste10}. As a result, recent studies have shifted to using dark matter halo masses as large as $10^{11}$~M$_{\odot}$ \citep{purcell11,gomez15} based on halo abundance matching arguments. Such high values are comparable to the mass of the LMC progenitor \citep[see e.g.][]{penarrubia16, jethwa16} and imply a mass ratio relative to the MW on the order of 1:10. However, unlike the Magellanic Clouds, which may be on their first passage near the MW \citep{besla07}, Sgr is known to have experienced multiple close passages in the past. If true, such high Sgr progenitor masses would have important implications for the formation and evolution of the MW disk \citep[e.g.][]{purcell11, gomez13, donghia16}. Because the dynamical friction force is proportional to the square of the satellite mass, we expect drag to play a much more important role in slowing down the Sgr satellite and bringing it to closer Galactocentric distances. Here we perform an exploration across orbital angular momentum parameter space analogous to \S~\ref{subsec:light_param_exploration}, this time using a Sgr progenitor mass of $6\times10^{10}$~M$_{\odot}$ according to the recent estimates of \citet{gibbons17}. We consider two possibilities: a `slow sinking' scenario in which, as in \S \ref{sec:light_Sgr}, Sgr crosses the MW virial radius at $z\sim1$ (approximately 8~Gyr ago); and a `rapid sinking' scenario, in which we examine a first infall around $z\sim0.4$, about 4~Gyr ago.

\subsection{Dynamical friction formalism}
\label{subsec:df_formalism}

\begin{figure*}[hbt!]
\begin{center}
\includegraphics[scale=0.5]{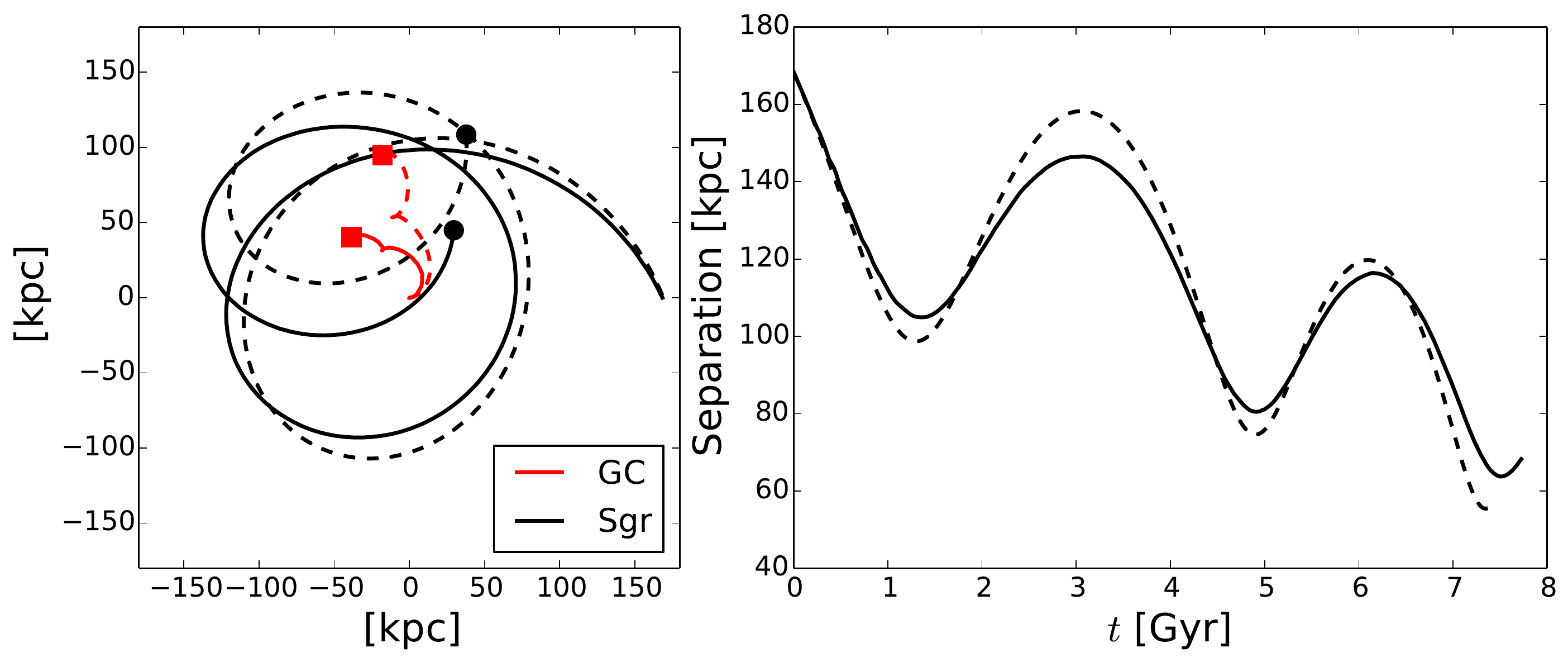} 
\caption{Comparison of an example Sgr orbit computed with GADGET and the semianalytic model after tuning of the dynamical friction described in \S~\ref{subsec:df_formalism} (solid and dashed lines, respectively). \textbf{Left panel:} trajectories of the Sgr progenitor (black) and MW Galactic Center (red) in the Sgr orbital plane. The current dynamical centers of the two galaxies are indicated by a colored dot and square, respectively. The MW barycenter is significantly displaced by the interaction with a massive Sgr, an effect previously illustrated by \citet{dierickx14} for Andromeda's satellite M32, and by \citet{gomez15} for the LMC. \textbf{Right panel:} separation between the two galaxies as a function of time since the beginning of the calculation.  
\label{fig:dftuning} }
\end{center}
\end{figure*}

With a Sgr progenitor mass of $6\times10^{10}$~M$_{\odot}$, the total mass ratio of the MW-Sgr merger is approximately 17:1. In this regime we expect the semianalytic model we have relied upon so far, tuned for a 100:1 mass ratio, to require renewed calibration. Because the total mass of the Sgr progenitor is no longer negligible compared to the MW mass interior to the Sgr orbit, Chandrasekhar's approximation is an inexact description of dynamical friction. The amount of stripped material can be significant for a massive satellite with close pericentric passages. Therefore the self-gravity of the wake, not taken into account in the classical Chandrasekhar formula, could begin to affect the drag. Additionally, the material stripped from the satellite becomes bound to the MW halo where it contributes to reshaping its structure. However, the Sgr mass outside the tidal radius is assumed for simplicity to vanish from the calculation in our model. With these caveats in mind, we conduct a simple series of semianalytic/$N$-body comparisons in order to tune dynamical friction iteratively. 

Chandrasekhar proposed the following form for the acceleration introduced by dynamical friction \citep[][eq. 7.18]{binney87}:
\begin{equation}
\vec{a}_{\text{DF}} = - \frac{4 \pi \ln(\Lambda) G^2 \rho(r) M_{\text{Sgr}}(< r_t) }{v^3} \times \sqb{\erf(X) - \frac{2X}{\sqrt{\pi}} \exp(-X^2) } \vec{v} \ ,
\end{equation}
Here $M_{\text{Sgr}}(< r_t)$ is the Sgr mass interior to its minimum tidal radius $r_t$, $\rho(r)$ is the density of the MW halo at Galactocentric distance $r$, $\ln(\Lambda)$ is the Coulomb logarithm, and $X$ is defined as $X = v/\sqrt{2}\sigma$, where $v$ is the satellite velocity and $\sigma$ is the one-dimensional velocity dispersion of particles in the host halo \citep[given by][eq. 10]{hernquist90}. As in Paper I, we adopt an alternative time-dependent Coulomb logarithm defined following \citet{hashimoto03}:
\begin{equation}
\ln (\Lambda) = \ln \bra{\frac{r}{1.4 \epsilon}} \ ,
\end{equation}
where $\epsilon$ is a softening length variable defined by the authors to model the LMC with a Plummer sphere. In adjusting the friction parameterization, we find that setting $\epsilon \simeq 2$~kpc in the semianalytic model of a massive Sgr yields the best agreement with $N$-body integrations of the same initial conditions. In Figure~\ref{fig:dftuning} we show the semianalytic model trajectory calculated with this parameterization superimposed on an analogous $N$-body orbit. Compared with $\epsilon \simeq 1$~kpc for the lower mass Sgr progenitor case analyzed in \S~\ref{sec:light_Sgr}, the new value is larger by a factor of 2, qualitatively in line with expectations considering a mass increase by a factor of 6. Our parameterization is also consistent with the larger value used by \citet{hashimoto03} for the more massive LMC.

\subsection{Slow sinking}
\label{subsec:massive_slow}

\begin{figure*}[hbt]
\begin{center}
\includegraphics[scale=0.27,trim = 0mm 0mm 0mm 0mm, clip]{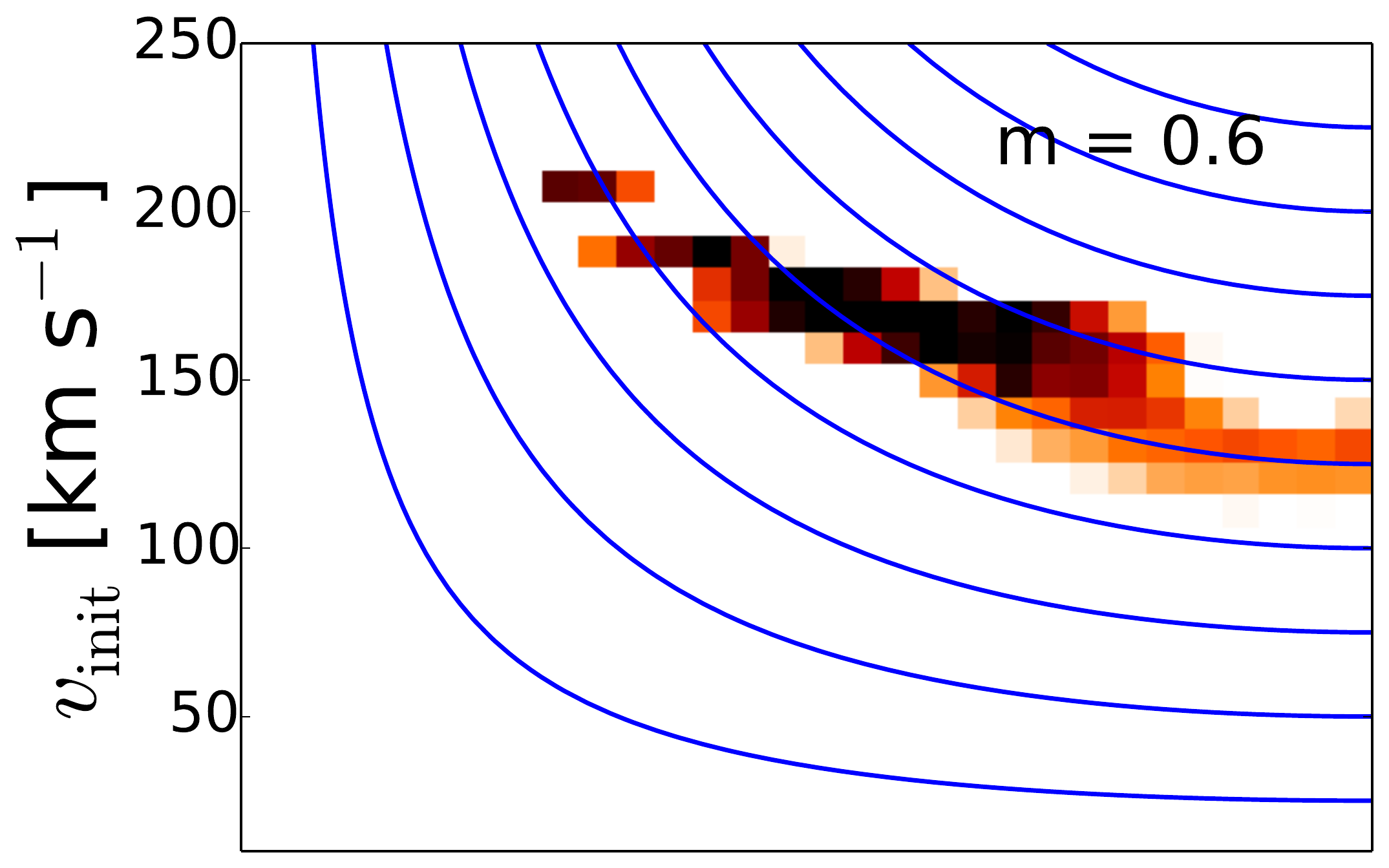} 
\includegraphics[scale=0.27, trim = 0mm 0mm 0mm 0mm, clip]{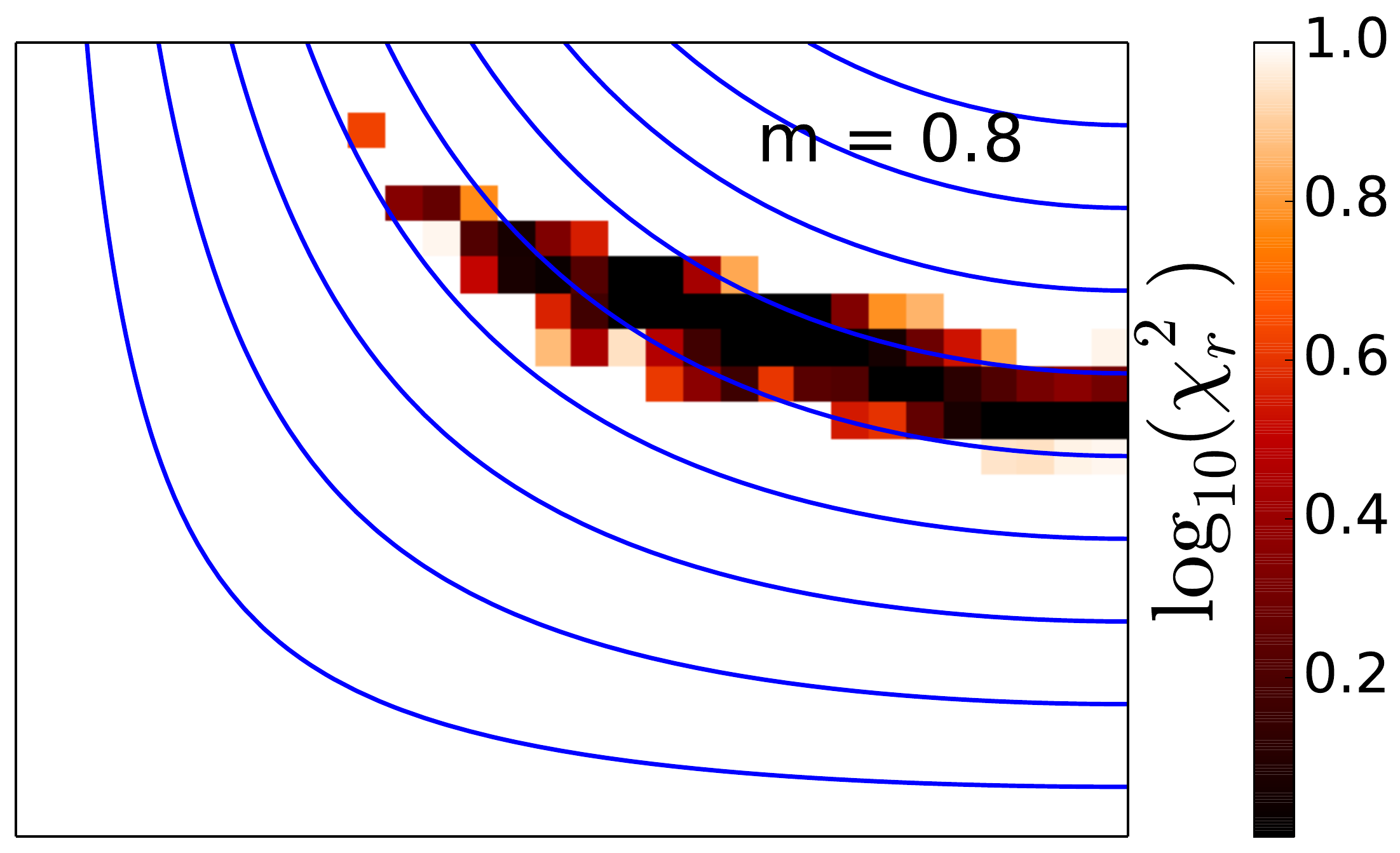} \\
\includegraphics[scale=0.27, trim = 0mm 0mm 0mm 0mm, clip]{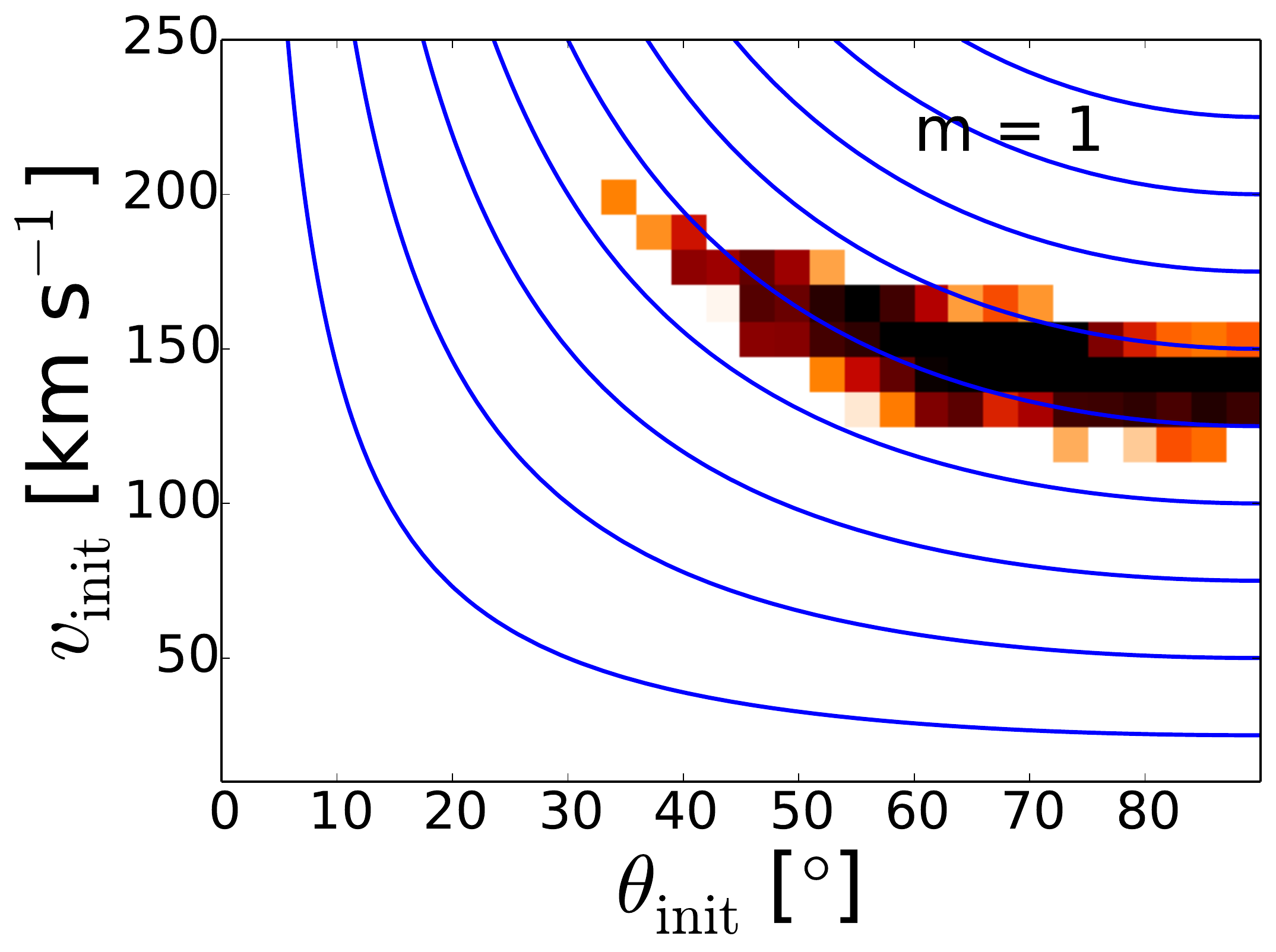} 
\includegraphics[scale=0.27, trim = 0mm 0mm 0mm 0mm, clip]{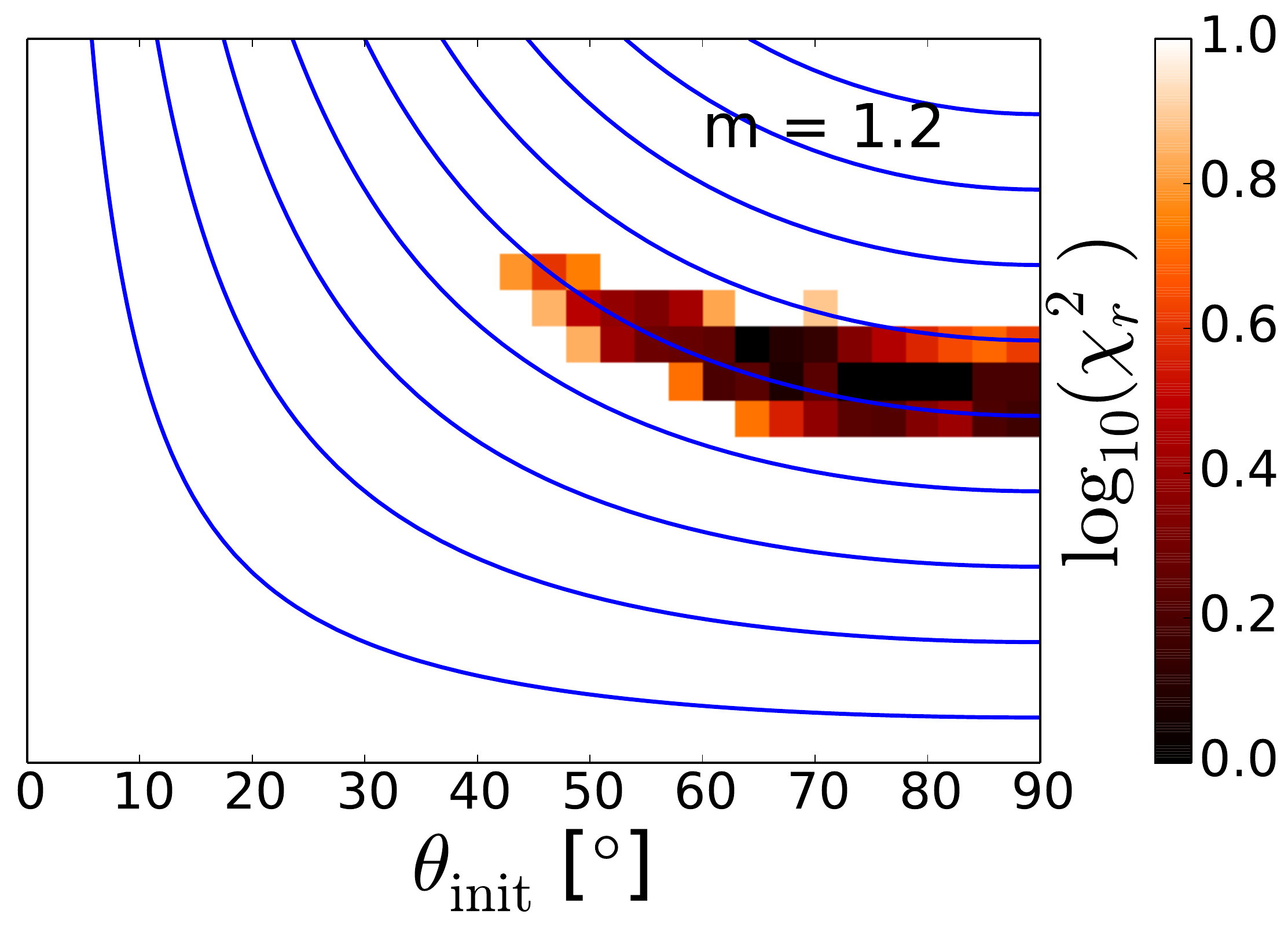} \\
\includegraphics[scale=0.3, trim = 0mm 0mm 0mm 2mm, clip]{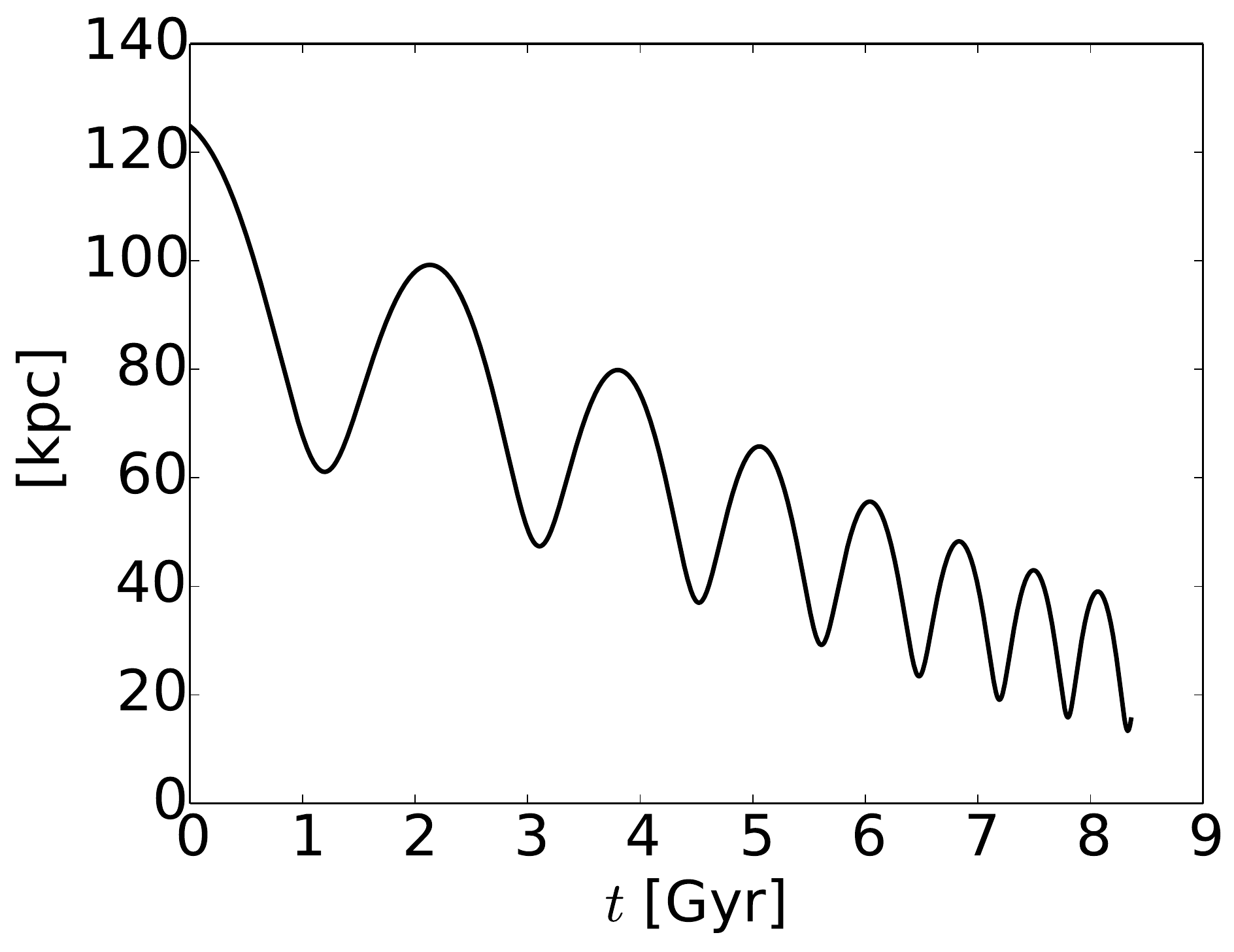} 
\includegraphics[scale=0.27, trim = 0mm 0mm 0mm 0mm, clip]{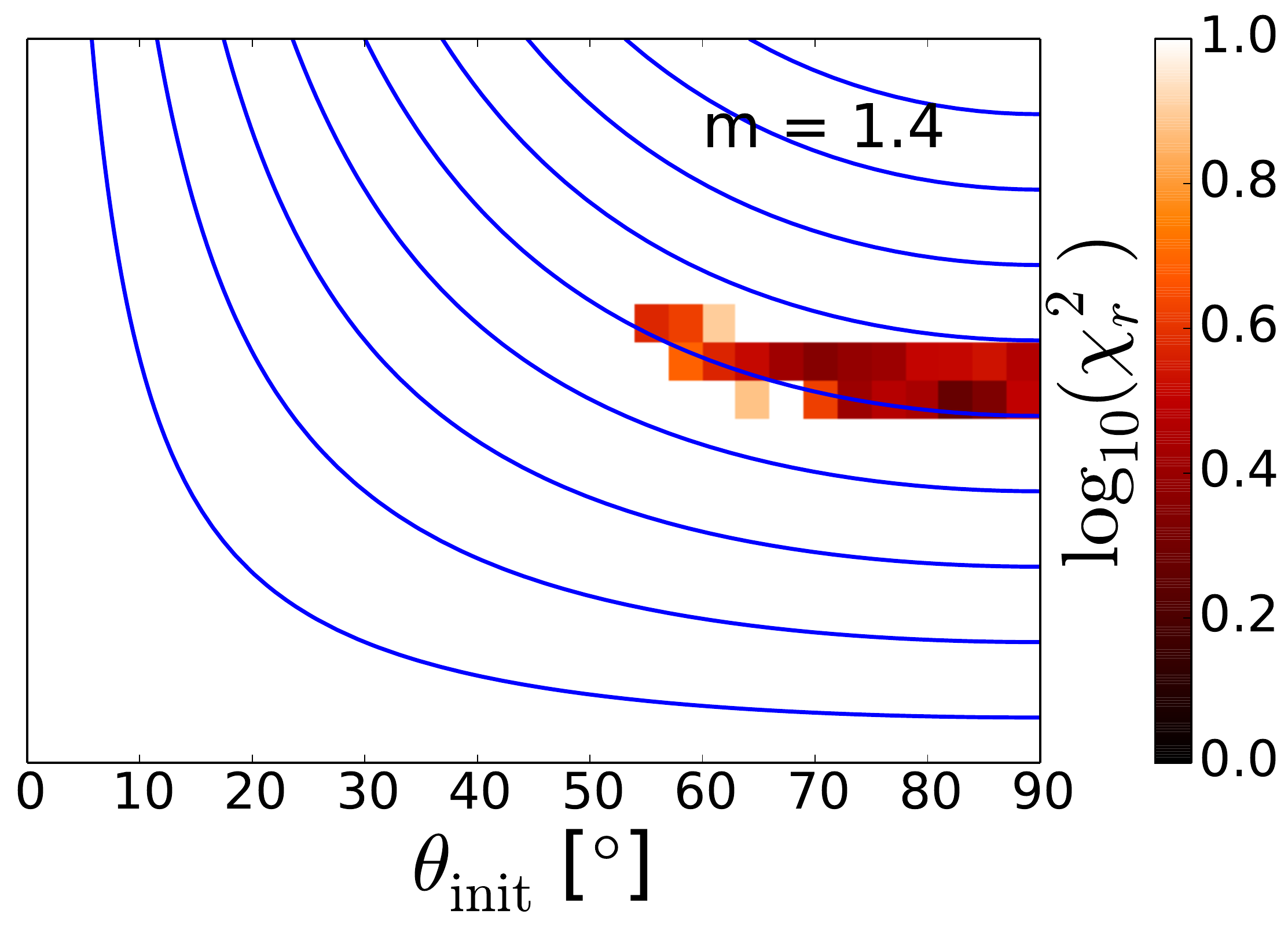} 
\caption{As in Fig~\ref{fig:ridge1}, color maps of match to Sgr phase-space coordinates for different MW masses, for the slow sinking scenario of a massive Sgr progenitor. The bottom left panel shows the separation between the MW and Sgr as a function of time, for the best-match orbit in the fiducial $m=1$ case. 
\label{fig:ridge2} }
\end{center}
\end{figure*}

As in \S~\ref{subsec:light_param_exploration}, Figure~\ref{fig:ridge2} presents color maps of the regions of parameter space consistent with the present-day position and velocity of Sgr, for five different values of the MW mass. Fewer matching orbits are found across the parameter space, and there appears to be a modest dependence on the MW mass. The few successful combinations occur closer to the high initial angle and high velocity corner of parameter space than in the low-mass case, at large values of the initial orbital angular momentum. This is to be expected given that in the framework considered here, the satellite galaxy is subjected to 7-9~Gyr of evolution under strong dynamical friction. Therefore a high initial orbital angular momentum is needed for the dwarf remnant to survive to the present day. The best fit orbit for the fiducial case $m=1$ is shown in the bottom left panel of Figure~\ref{fig:ridge2} as an example. We note the gradual sinking of the orbit to lower pericentric distances under the effect of dynamical friction. 


The requirement of matching Sgr phase-space coordinates appears less sensitive to variations in the MW mass than in the low-mass Sgr progenitor case. This occurs because dynamical friction is now sufficiently effective to modulate the trajectory, leading to orbital solutions across the full range of host masses. However, the type of orbits that are allowed varies across the MW mass range we explore. As the MW mass increases, the preferred region of parameter space moves from initial velocities $\simeq150-180$~km~s$^{-1}$ and angles $<65^\circ$, to lower $v_\text{init}$ ranges of $\simeq130-150$~km~s$^{-1}$ and higher $\theta_\text{init}$ in the $65-90^\circ$ range. For a low MW mass, the Sgr orbit is essentially required to be more eccentric, while in the high-mass case the infall starts off in a more circular manner.

Close to circular orbits are improbable from a statistical perspective in the cosmological galaxy formation paradigm. The modest spin parameters of mature dark matter haloes imply an upper limit on the amount of angular momentum gained through mergers. Because the eventual halo rotation is small, most satellites must fall in on primarily radial orbits. In the simulated satellite population of \citet{wetzel10}, the average ratio of radial to tangential velocity components is 1.4. With initial velocity vectors almost purely in the tangential direction, the Sgr orbits found for the higher MW masses in Figure~\ref{fig:ridge2} present a reversed ratio and would strongly deviate from the expectation for satellite infall models.

\subsection{Rapid sinking}
\label{subsec:massive_rapid}

\begin{figure*}[hbt]
\begin{center}
\includegraphics[scale=0.27,trim = 0mm 0mm 0mm 0mm, clip]{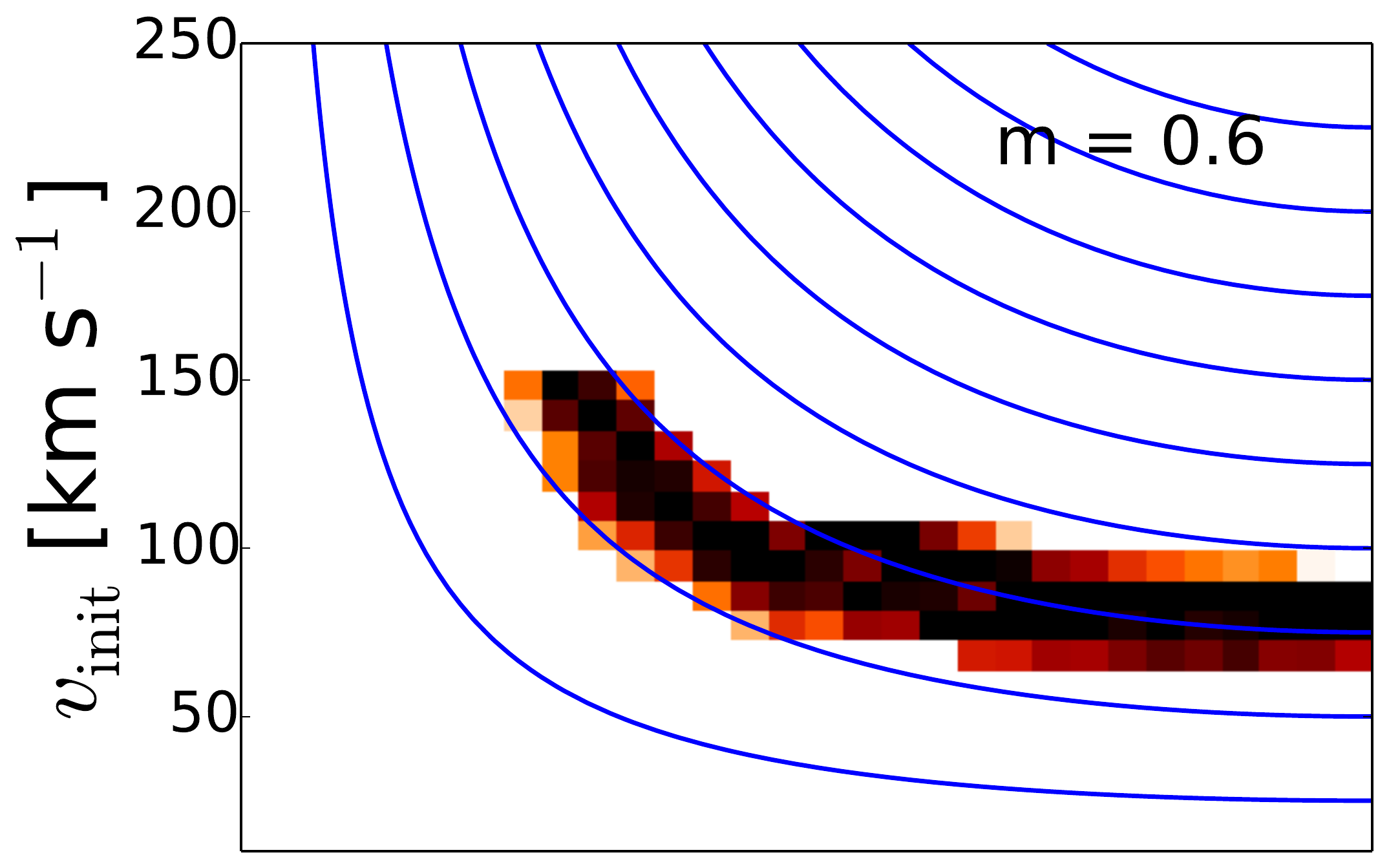} 
\includegraphics[scale=0.27, trim = 0mm 0mm 0mm 0mm, clip]{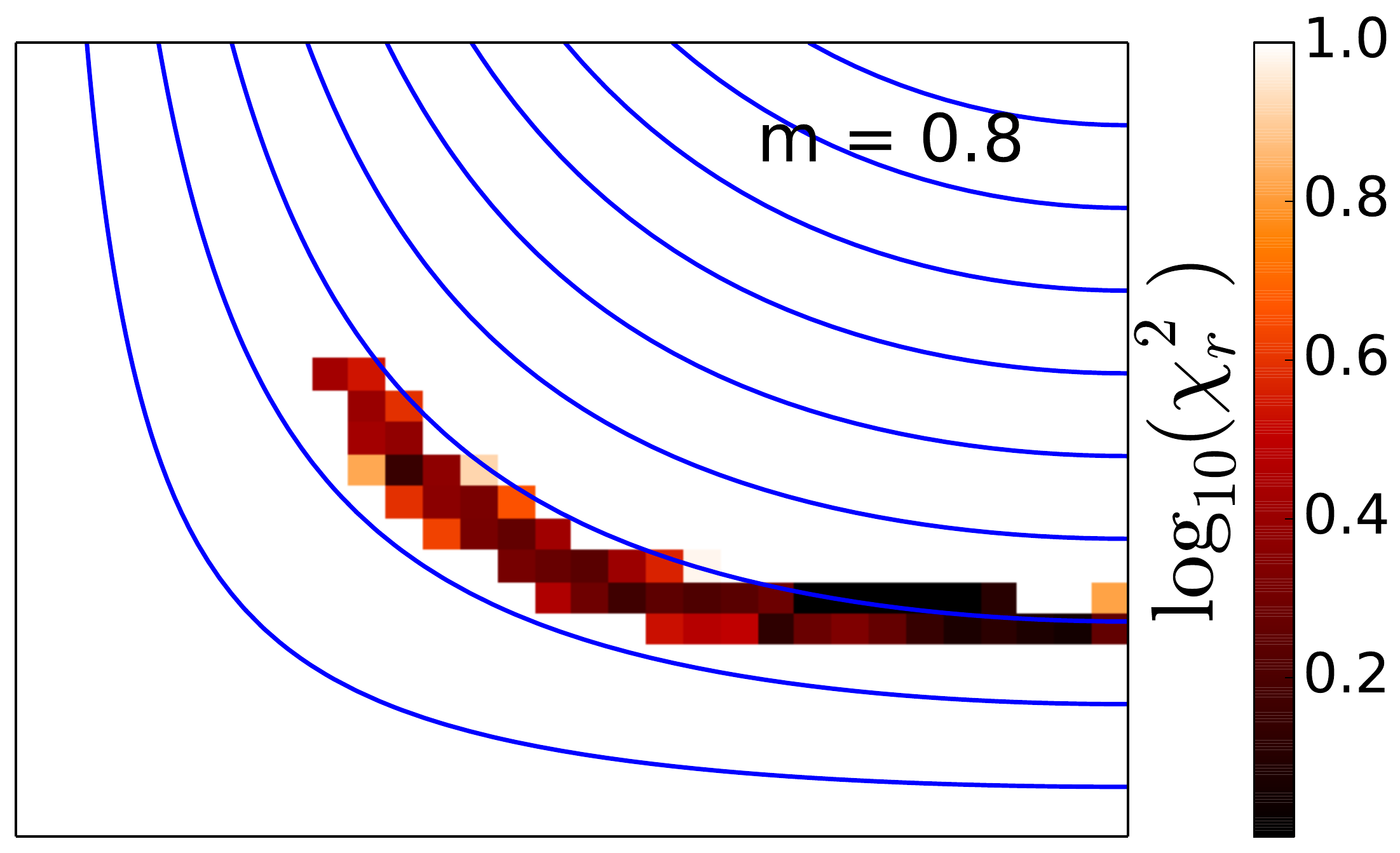} \\
\includegraphics[scale=0.27, trim = 0mm 0mm 0mm 0mm, clip]{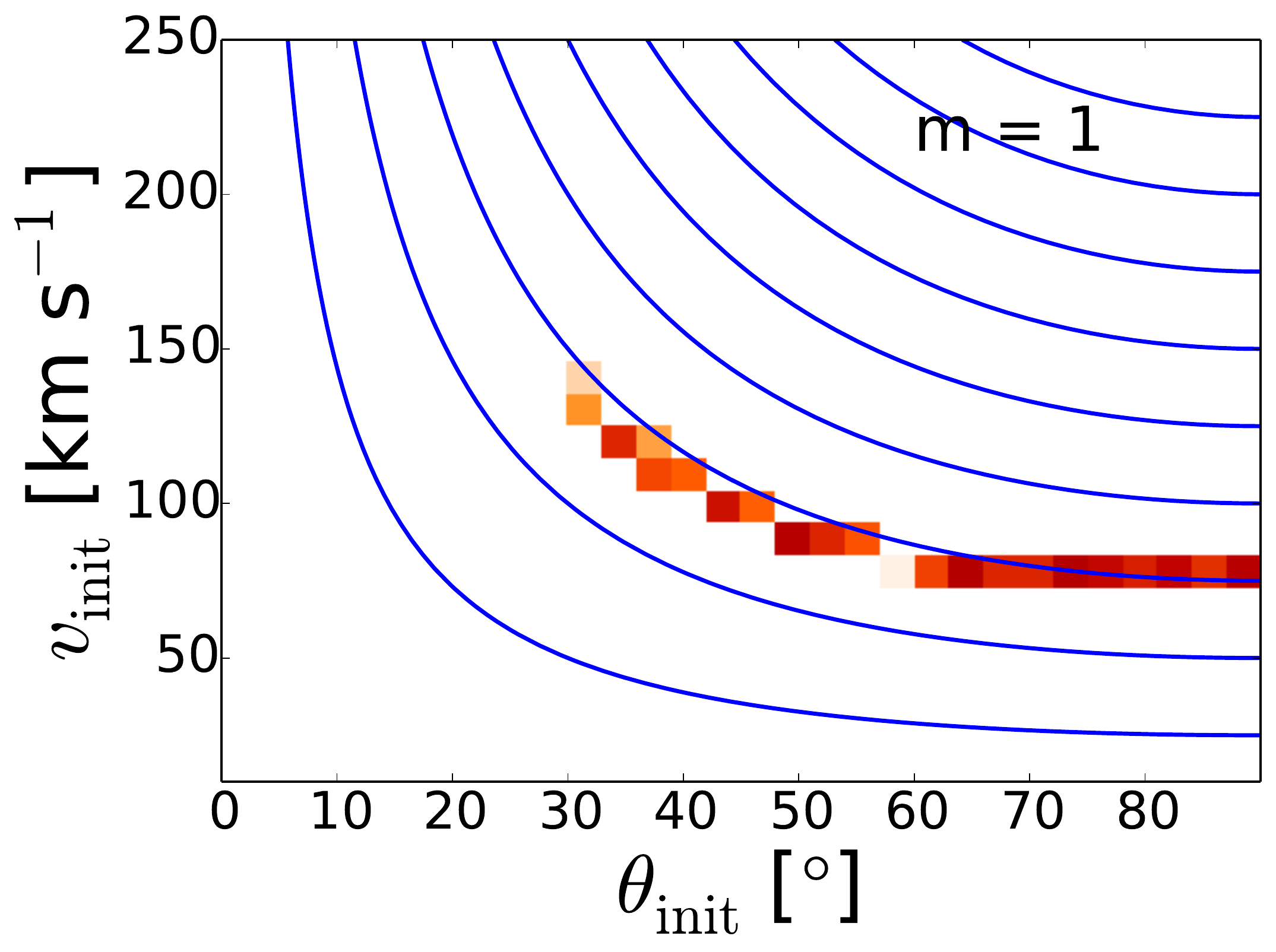} 
\includegraphics[scale=0.27, trim = 0mm 0mm 0mm 0mm, clip]{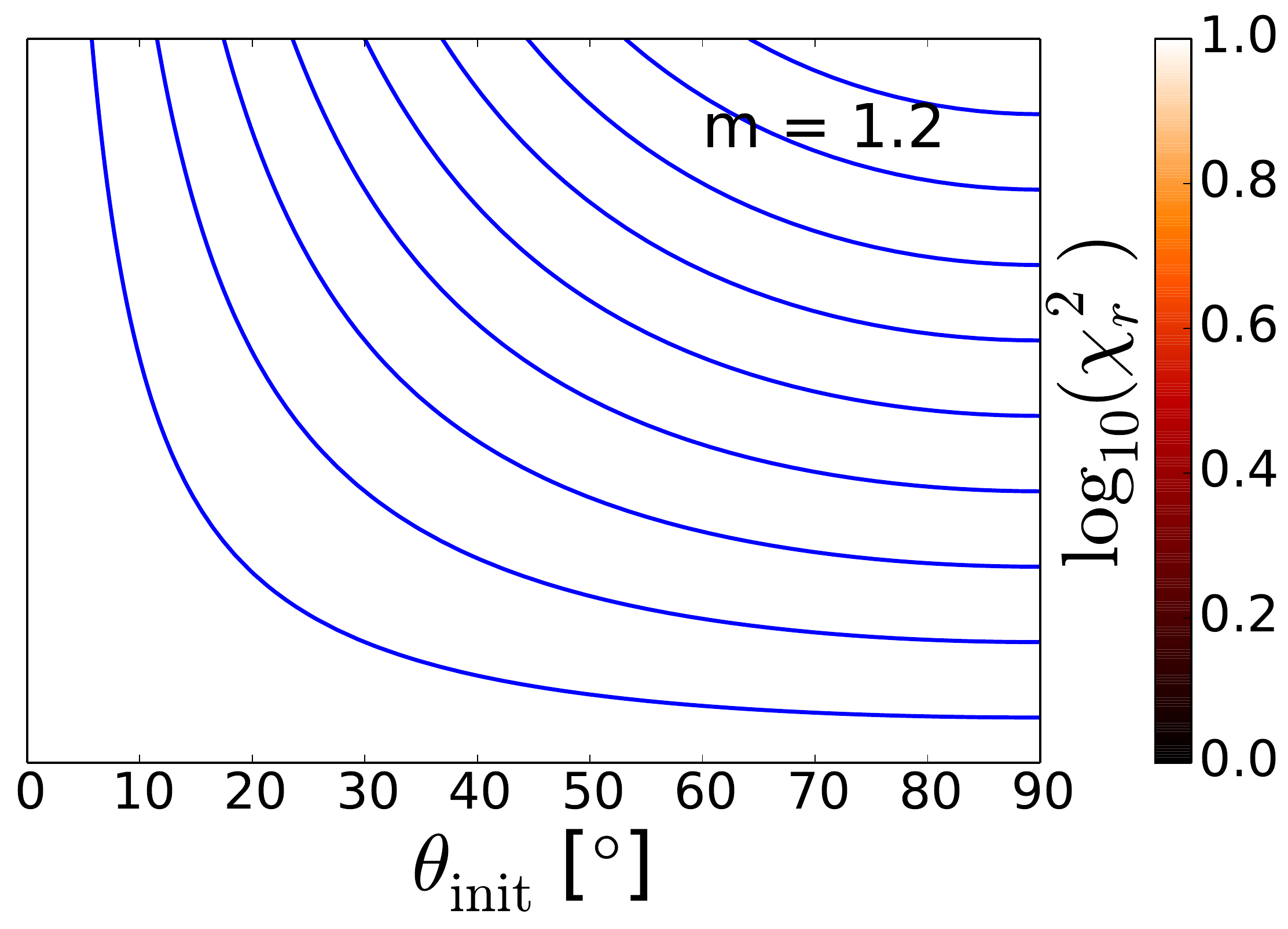} \\
\includegraphics[scale=0.3, trim = 0mm 0mm 2mm 0mm, clip]{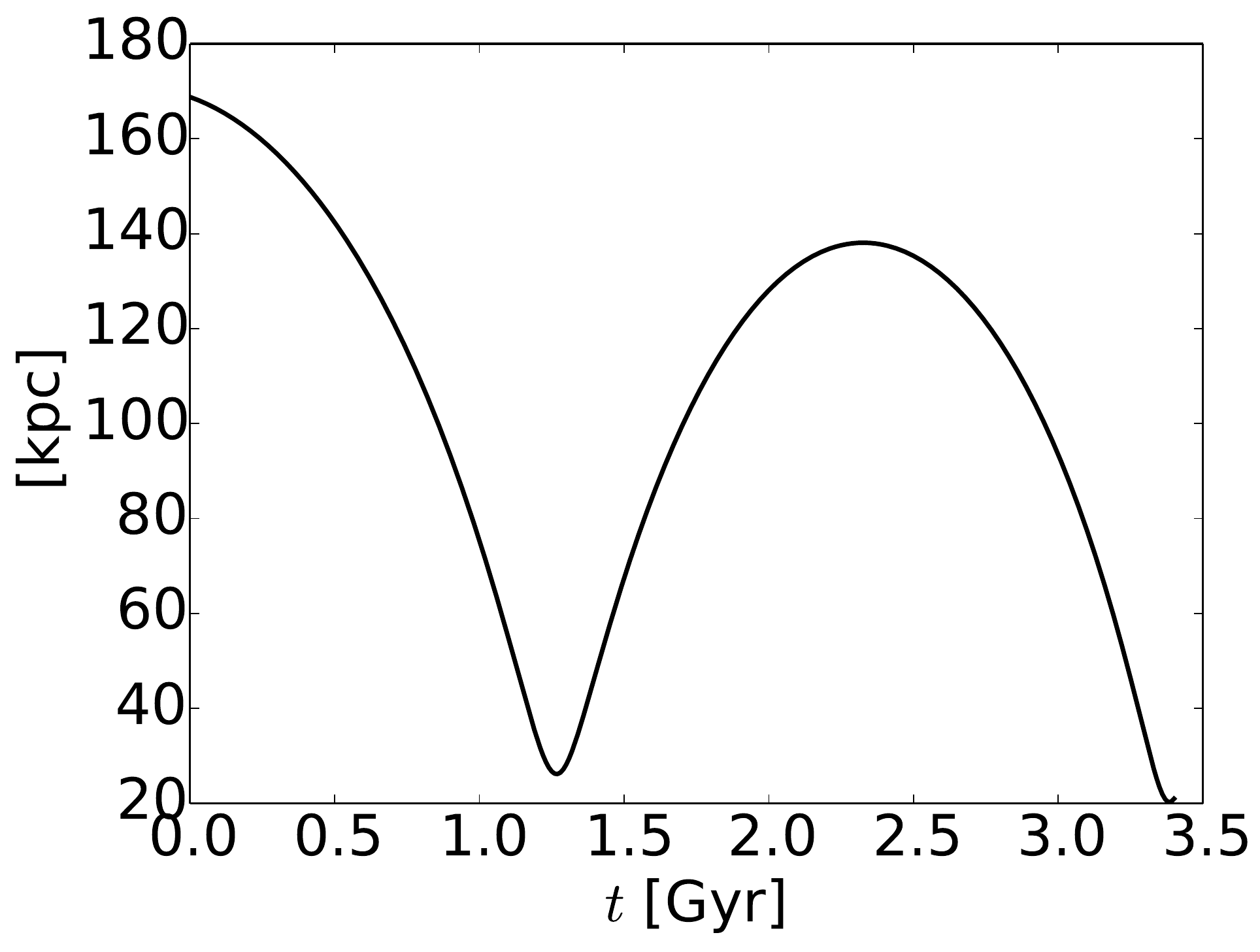} 
\includegraphics[scale=0.27, trim = 0mm 0mm 0mm 0mm, clip]{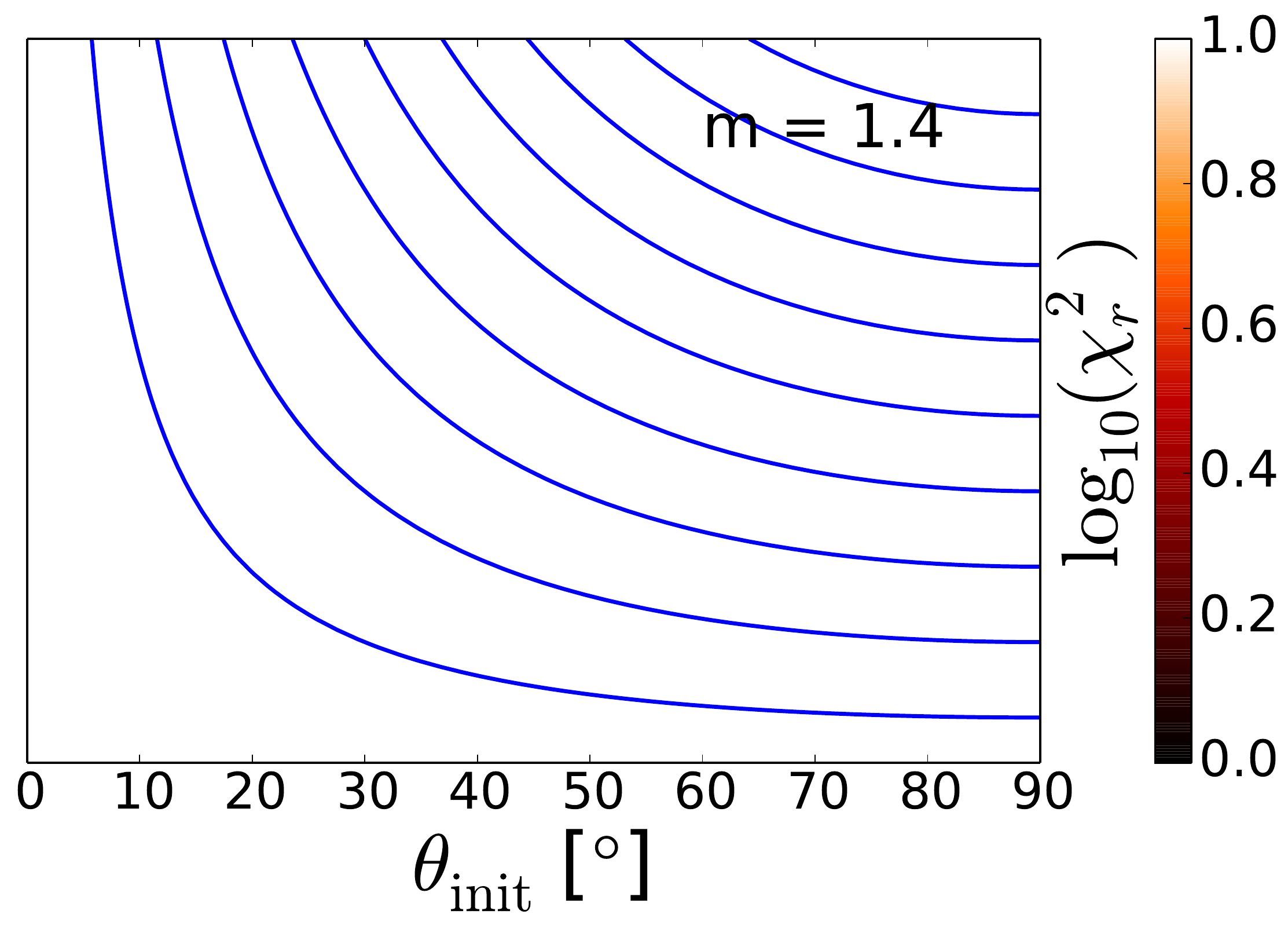} 
\caption{As in Figs~\ref{fig:ridge1} and \ref{fig:ridge2}, color maps of match to Sgr phase-space coordinates for different MW masses, in the rapid sinking scenario of a massive Sgr progenitor. 
\label{fig:ridge3} }
\end{center}
\end{figure*}

The results of \S~\ref{subsec:massive_slow} rely on the framework we have adopted so far, where the starting conditions for the Sgr infall are a virial crossing redshift of $z=1$ with an initial separation corresponding to the MW virial radius at that redshift. Working with these assumptions, we found few satisfactory trajectories for a high Sgr progenitor mass, and a modest dependence on the MW virial mass. Here we investigate whether successful orbits can be recovered if the initial conditions are amended to reflect faster infall due to stronger dynamical friction. We test this possibility with an orbital integration of 4~Gyr, half of the previously considered 8~Gyr of evolution. Sgr initial distances are chosen to match the MW virial radius at that time for each different MW model, as summarized in Table~\ref{tab:params}. For the fiducial $m=1$ case, the virial radius is calculated assuming a virial mass at $z\sim0.37$ of approximately $8\times10^{11}$~M$_{\odot}$, following cosmological galaxy formation simulations \citep{torrey15, lu16}. The Sgr starting distances corresponding to the MW virial radii for the four other cases are calculated by proportionally scaling the fiducial mass. 

Figure~\ref{fig:ridge3} shows the regions of parameter space that lead to matches with the present-day coordinates of the Sgr remnant. Remarkably, we recover the same behavior found earlier in \S~\ref{subsec:light_param_exploration} for an initial Sgr mass lower by a factor of 6. A similar narrow strip of parameter space is favored across different values of the MW mass. The preferred range of initial orbital angular momenta is the same as in the lower-mass case (see Figure~\ref{fig:ridge1}). Furthermore, as the MW mass increases, we find an even steeper decline in the area of parameter space compatible with present-day constraints. This indicates that an infall epoch as recent as $z\sim0.4$ is implausible if both the MW and Sgr haloes are on the high end of their putative mass ranges, in agreement with expectations from cosmological simulations. Again we provide the best fit orbit for the fiducial case $m=1$ as an example in the bottom left panel. We point out that the quality of satisfactory orbits in the $m=1$ case is marginal, and that for the best-match orbit in this framework Sgr has only undergone one pericenter passage prior to the present day. One passage at low Galactocentric distances may not be sufficient to give rise to streams similar to the rich structure observed in reality. This serves as a further illustration that in a rapid infall scenario for a massive Sgr progenitor, MW masses $<10^{12}~M_{\odot}$ are preferred.

\section{Discussion and Conclusions}
\label{sec:conclusions}

Dynamical modeling by \citet{jiang00} demonstrated that a wide range of orbital histories for Sgr are possible depending on the initial mass of the progenitor. Ranging over two orders of magnitude from $10^9$ to $10^{11}$~M$_{\odot}$, the progenitor mass is tightly coupled to the initial apocentric distance, a dependence mediated by dynamical friction. Bracketing the range of $(1-6)\times10^{10}$~M$_{\odot}$, we have tested the orbital dynamics of Sgr for different MW host models. We have restricted the analysis to this range for two main reasons:
\begin{itemize}
\item \textit{Plausibility of close-in formation:} As exemplified by the models of \citet{law10}, low Sgr masses imply the progenitor experienced little to no dynamical friction, and therefore originated at distances comparable to the present-day observed stream apocenters (50-100~kpc). This implies that it either formed deep inside the MW halo, conflicting with simulations of galaxy formation, or early on in the MW's history when its virial radius was much smaller. Such early accretion is implausible given the Sgr age derived from stellar population studies, which estimate the beginning of disruption at approximately 5-8~Gyr ago \citep{bellazzini06, deboer15}.
\item \textit{Observational evidence for a higher Sgr mass:} Recent detailed studies of the stellar content of the Sgr remnant and tidal stream have yielded large mass estimates. The best-fit models to the stellar dynamics of the Sgr core observed with APOGEE have total dark matter masses in the range of $(5-8)\times10^8$~M$_{\odot}$ \citep{majewski13}. The initial mass must have been at least one order of magnitude larger in order to account for the mass lost to tidal stripping. This is consistent with the values estimated from abundance matching using the luminosity tally of \citet{niederste10} \citep{conroy09, behroozi10, garrison-kimmel17}.
\end{itemize}

We find that in the low-mass case (M$_\text{Sgr, init} = 10^{10}$~M$_\odot$), the condition of attaining the correct present-day Sgr phase-space coordinates favors low MW masses ($\leq 10^{12}$~M$_\odot$). Extending our analysis to the high-mass end of the allowed range for Sgr (M$_\text{Sgr, init} = 6\times10^{10}$~M$_\odot$), we tested two different sets of assumptions regarding the satellite's initial apocentric distance and virial radius crossing epoch. We find that in the slow infall scenario, high initial orbital angular momentum is needed in order to counteract strong dynamical friction and ensure the satellite's survival to the present day. In this framework, a high MW mass is still disfavored by the model, partly because it would require a cosmologically improbable more circular orbit for Sgr. In a faster infall scenario, where Sgr crossed the MW virial radius at $z\sim0.4$, the constraints on the MW mass and the initial orbital angular momentum of the Sgr progenitor are even stronger than in the low-mass case.

The orbit of Sgr can be seen as a clock measuring the MW gravitational potential. \citet{penarrubia06} have argued that the properties of tidal streams reflect the present-day Galactic potential only. While it appears that stream models have little constraining power on the past evolution of the MW, the satellite orbits giving rise to the stream are sensitive to the time dependence of the host potential \citep{knebe05}. Given the initial conditions of our model and assumption of a static MW potential, our analysis yields an upper limit on the mass inside $\sim125$~kpc at redshift $z\sim1$. We show that high masses in the inner regions of the MW at $z=1$ are inconsistent with the present-day observed Sgr phase-space coordinates. Additional mass since $z=1$ would worsen the discrepancy, strengthening our estimated upper limit. 

Our finding of a lightweight Galaxy based on matching the coordinates of the Sgr remnant is qualitatively consistent with a previous estimate by \citet{gibbons14} derived from stellar stream properties. The fact that both studies have independently come to similar conclusions by analyzing separate facets of the Sgr system lends weight to existing evidence favoring a lower Galactic mass \citep[e.g.][]{battaglia05, xue08, deason12a, bovy12, rashkov13, williams15, williams17}. This finding has important implications for the $\Lambda$CDM paradigm of galaxy formation. A lower mass MW helps to alleviate the so-called Missing Satellites \citep{klypin99} and Too Big Too Fail problems (\citet{boylan-kolchin11}; see also \citet{oman16, lovell16, sawala17}). However, extremely low MW masses, such as that estimated by \citet{gibbons14}, may be in tension with the MW's stellar content \citep{taylor16}. A virial mass of at least $\sim8\times10^{11}$~M$_\odot$ is needed in order to reconcile the baryonic mass of the MW with the well-measured cosmological baryon matter fraction \citep{zaritsky17}. Such lines of investigation are especially valuable because they provide independent evidence from the mass estimates above derived based on dynamical methods. With new detections of the Sgr stellar stream, future data from {\it Gaia}\footnote{http://sci.esa.int/gaia/} and the Wide Field Infrared Survey Telescope\footnote{http://wfirst.gsfc.nasa.gov} (WFIRST) will allow to improve the modeling of the Sgr orbit presented in this paper. 

\acknowledgments
We thank Laura Blecha for helpful discussions.

\acknowledgments

\end{document}